\documentclass[fleqn,usenatbib,twocolumn]{mnras}
\usepackage[utf8]{inputenc}
\usepackage{graphicx}
\usepackage{amsmath}
\usepackage{mathtools}
\usepackage{siunitx}
\usepackage{enumerate}
\usepackage{caption,subcaption}
\usepackage{comment}
\usepackage{animate}

\newcommand{\acen}{$\alpha\;$Cen}
\newcommand{\tav}{t_{\rm av}}
\let\vec=\mathbfit
\newcommand{\vu}{\vec{u}}
\newcommand{\vun}{\hat{\vec{u}}_0}
\newcommand{\gw}{g_{\scriptscriptstyle W}}
\newcommand{\ri}{\vec{X}_i}
\newcommand{\rj}{\vec{X}_j}
\newcommand{\vi}{\vec{V}_{\!i}}

\newcommand{\cubed}{^{\rlap{$\scriptstyle3$}}}
\newcommand{\half}{{\textstyle\frac12}}
\newcommand{\rtide}{r_{\rm tid}}
\newcommand{\rab}{r_{\scriptscriptstyle\!AB}}
\newcommand{\Xwob}{\widetilde X}
\newcommand{\Ywob}{\widetilde Y}

\title{Simulations of astrometric planet detection in Alpha~Centauri by
intensity interferometry}

\author[Rai et al.]{Km Nitu Rai,$^{1}$\thanks{E-mail: niturai20129617@iisertvm.ac.in}
Subrata Sarangi,$^{2}$ Prasenjit Saha$^{3}$\thanks{E-mail: psaha@physik.uzh.ch} and Soumen Basak$^{1}$\\ \\
$^{1}$School of Physics,
Indian Institute of Science Education and Research Thiruvananthapuram, Maruthamala PO,
Vithura,\\ Thiruvananthapuram 695551, Kerala, India \\
$^{2}$School of Applied Sciences, Centurion University of Technology and Management, Odisha-752050, India \\
$^{3}$Physik-Institut, University of Zurich, Winterthurerstrasse 190, 8057 Zurich, Switzerland}

\date{}

\begin{document}

\maketitle

\begin{abstract}
Recent dynamical studies indicate that the possibility of an
Earth-like planet around \acen~A or B should be taken seriously.  Such
a planet, if it exists, would perturb the orbital astrometry by
$<\SI{10}{\micro{\rm as}}$, which is $10^{-6}$ of the separation
between the two stars.  We assess the feasibility of detecting such
perturbations using ground-based intensity interferometry.  We
simulate a dedicated setup consisting of four 40-cm telescopes
equipped with photon counters and correlators with time resolution
$0.1\,\rm ns$, and a sort of matched filter implemented through an
aperture mask.  The astrometric error from one night of observing
\acen~AB is $\approx0.5\,\rm mas$.  The error decreases if longer
observing times and multiple spectral channels are used, as
$(\hbox{channels}\times\hbox{nights})^{-1/2}$.
\end{abstract}


\section{Introduction}

Detecting (or ruling out) a habitable planet around one of
$\alpha$~Centauri A or B is currently just a little beyond available
techniques, and thus remains a very intriguing possibility.  The two
bright stars in \acen\ are an 80\thinspace yr binary with masses
$1.11M_\odot$ and $0.94M_\odot$. The system also has a third star
(Proxima Centauri) of mass $0.12M_\odot$ much further away \citep[the
  orbital period is about 0.5~Myr, see e.g.,][]{2017A&A...598L...7K}.
Proxima actually has an Earth-mass planet in the nominal habitable
zone, inferred from a radial velocity perturbation of
$\sim\SI{1}{\metre\sec^{-1}}$ \citep{2016Natur.536..437A}, and a
further planet candidate \citep{2020SciA....6.7467D}. However, the
eruptive brightness changes of the star makes the Proxima~b planet
less promising as habitable. Hence there is great interest in a
possible Earth-mass planet with an orbit $\sim1\,$au
around $\alpha$~Centauri A or B.  For example, see the composition
model by \cite{2021arXiv211012565W}.

Orbit integrations by \cite{1988A&A...206..143B} and more recently by
\cite{2018AJ....155..130Q} indicate that stable orbits can exist in
the habitable zones of \acen~A and B.  Rocky planets could be present
in such an orbit, and be undetectable with present or planned
instruments.  Inclinations that produce transits from our line of
sight are unlikely.  The radial-velocity amplitude for an Earth-mass
planet would be only $\sim\SI{10}{\centi\metre\sec^{-1}}$.
Coronographic imaging by the James Webb Space Telescope is an exciting
prospect, but the smallest detectable radius is estimated to be
$3R_\oplus$ \citep{2020PASP..132a5002B}.  A further possible detection
path is astrometry of the host star.  Optical interferometry of the
binary HD~176051 reveals a planet through perturbations of the
relative astrometry of the two stars
\citep{2010AJ....140.1657M}. Perturbation in absolute astrometry,
measured using radio interferometry, revealed a planet in TVLM
513-46546 \citep{2020AJ....160...97C}. Gaia astrometry is expected
over time to produce many exoplanet discoveries
\citep{2014ApJ...797...14P}.  However, none of these can reach
$<\SI{10}{\micro{\rm as}}$ level of an Earth-mass planet 1\thinspace
au from \acen~A or B.  Relative astrometry of the binary via optical
interferometry \citep[as applied to GJ65 by the GRAVITY
  collaboration:][]{2017A&A...602A..94G} is a possibility.  A
$\SI{30}{\centi\metre}$ space telescope optimised for astrometry could
reach $0.2\,\mu\rm as$ (or 1\thinspace picoradian), which would easily
be enough but such an instrument is so far only at the concept stage
\citep[TOLIMAN][]{2018SPIE10701E..1JT,2021AAS...23731802B}.

In this paper, we propose intensity interferometry as a novel way of
planet-searching in \acen~A,B. Astrometry of a binary system through
intensity interferometry goes back to \cite{1971MNRAS.151..161H}. The
\acen\ system, however, presents an additional challenge, because the
stars are far apart. The underlying interference pattern consists of
superposed Airy discs of radii $\sim\SI{10}{\metre}$ due to \acen\ A
and B, modulated by fringes of cm-scale width, the exact width varying
with the orbital epoch.  The large light buckets essential for
intensity would wash out the fringes. To deal with this challenge, we
suggest a special kind of masked aperture, which has a particular
masking strips arrangement, resulting in many baselines for a single
baseline.\footnote{This is unrelated to aperture masking or
diffractive pupils in imaging telescopes
\cite[e.g.,][]{2020SPIE11446E..11S}, which are about introducing
diffraction features into an image for calibration.} To validate the
methodology described in this work, we simulate a dedicated setup of
four masked $\SI{40}{\centi\metre}$ telescopes observing \acen~A,B
from a circumpolar location such as Mt.~John NZ.

The paper is organized as follows. In Section~\ref{sec:fringes}
we introduce the interferometric signal, describe a simulated telescope
layout with aperture masks, and explain the signal-to-noise ratio (SNR).
This is followed by the Section~\ref{sec:orbits} where we discuss
about the orbits of a four-body system under consideration. In
Section~\ref{sec:Results} we discuss the results, the recovery of the
relative sky locations of \acen~A and B, using dynamic nesting
\citep{skilling2004nested,speagle2020dynesty} in particular, and verify
that the error has the expected dependence on the total SNR. In addition
to this, we also show how the $80\,$yr and $\sim1\,$yr proper-motion
components can be disentangled. Comparing the $\sim1\,$yr variation with the
astrometric accuracy we then estimate the detection threshold. Finally in
Section~\ref{sec:discussion} we discuss the prospects for
such an observation program and conclude.

\section{The Signal}\label{sec:fringes}

In this section we will recall the well-known interferometric signal
of a stellar binary, explain the difficulty caused by the wide
separation of stars, and introduce the idea of a matched aperture mask to
overcome it.  We will also derive the expected signal to noise ratio in
terms of instrument parameters.

\subsection{The interference pattern}

A featureless star of radius $R$ at distance $D$ has interferometric
visibility
\begin{equation}
  V(\rho) = 2 J_1(\eta)/\eta  \quad \hbox{where}\ \eta = 2\pi \rho R/D
\end{equation}
and $\rho$ is the radial coordinate in the usual interferometric
$(u,v)$ plane. In a realistic situation, Limb darkening (\cite{1970AJ.....75..175K, 2003A&A...412..241C}) of the star would modify the expression.

For a binary system, taking $X,Y$ as the physical separation of the
stars perpendicular to the line of sight, the visibility becomes
\begin{equation}
  V(u,v) =
  \frac{I_A V_A(\rho) + I_B V_B(\rho) \exp\left(2\pi i\,(uX + vY)/D\right)}
       {I_A + I_B}
\label{eqn:cvisib}
\end{equation}
where $I_A$ and $I_B$ are the brightnesses of the stars A and B. By
construction, $V(\rho=0)=1$. We will not consider a third star here,
as \acen~C is far-enough away to be easily excluded by the telescope.

Here, we are concerned with inferring the sky-projected
relative position $(X,Y)$ of the stars, assuming the distance is
known. It is straightforward to include additional parameters to be
fitted to data (the radii of both stars, parameters of a
limb-darkening model) as done in
\citep[cf.][]{10.1093/mnras/stab2391}. Such an approach is suitable
for close binaries. On the other hand, in case of \acen\, the stars are
far-enough apart to study individually to determine their radii and
limb-darkening coefficients. This has been done using Michelson interferometry by
\cite{2017A&A...597A.137K}, and in principle should be possible using
intensity interferometry as well. However, for this work we assume
the stellar radii are known, and disregard limb darkening.

In intensity interferometry the visibility $V(u,v)$ is a complex number and hence,
cannot be observed. Instead the square of magnitude of the visibility $V(u,v)$
\begin{equation}
  g(\vu) = |V(\vu)|^2 \quad \hbox{where}\ \vu \equiv (u,v)
  \label{eqn:gbasic}
\end{equation}
is observable as the excess intensity correlation.  This expression,
upon substitution of Eq.~(\ref{eqn:cvisib}) is equivalent to Eq.~(1) in
\cite{1971MNRAS.151..161H}.

As the Earth rotates, a pair of telescopes fixed on the ground will
trace a path in $u,v$ and the perpendicular coordinate $w$ along the
line of sight to the source. If the separation on the ground is
$(\Delta_{\scriptscriptstyle\rm E},\Delta_{\scriptscriptstyle\rm
  N},\Delta_{\rm up})$ then $u,v,w$ are given by the slightly
complicated rotation
\begin{equation}\label{eq:uvw-rotation}
\begin{pmatrix} u \\ v \\ w \end{pmatrix} = \frac1\lambda
R_1(\delta) \, R_2(h) \, R_1(-l)
\begin{pmatrix}
\Delta_{\scriptscriptstyle{\rm E}} \\
\Delta_{\scriptscriptstyle{\rm N}} \\
\Delta_{\rm up}
\end{pmatrix}
\end{equation}
where $l$ is the latitude of the setup, $\delta$ is the declination
and $h$ is the hour angle of the source, while
\begin{equation}
R_1(\delta) =
\begin{pmatrix}
1 & 0 & 0 \\
0 & \cos\delta  & -\sin\delta \\
0 & \sin\delta  &  \cos\delta
\end{pmatrix}
\label{eqn:R1}
\end{equation}
and similarly $R_1(-l)$
\begin{equation}
R_2(h) =
\begin{pmatrix}
 \cos h & 0 & \sin h \\
0       & 1 & 0 \\
-\sin h & 0 & \cos h
\end{pmatrix}
\label{eqn:R2}
\end{equation}
are the usual rotation matrices.  The $w$ coordinate does not change
the interference pattern, but it produces a time delay of $\lambda w/c$.

Starspots in either or both of the binary stars present an unknown
source of astrometric noise. \cite{catanzarite2008astrometric}
estimate that the effect would not severely affect astrometric
planet detection at Earth masses.  The largest sunspots would produce
a detectable astrometric signal \citep[cf.][]{hatzes2002starspots} but
the effect would have a very different time dependence than orbital
perturbations. Therefore, it is ignored in this work. 

\subsection{A masked aperture}

The signal \eqref{eqn:gbasic} applies if
\begin{enumerate}[(a)]
\item the time resolution (say $\Delta t$) is much finer than the
  coherence time (say $\Delta\tau$), and
\item the interference pattern does not vary significantly over
  the light bucket.
\end{enumerate}
Of these, (a) is never true in practice, with $\Delta t$ being orders
of magnitude more than the $\Delta\tau$.  As a result, the observed
signal is reduced by a factor of $\Delta\tau/\Delta t$. This effect
is well understood and always accounted for in observations. The condition
(b), on the other hand, does hold in previous applications, but not
for \acen, because the fringes are only a few centimetres apart.
Hence we have to account for the finite aperture of the light bucket.
\begin{figure*}
	\centering
	\begin{subfigure}{\linewidth}
		\includegraphics[width=\linewidth]{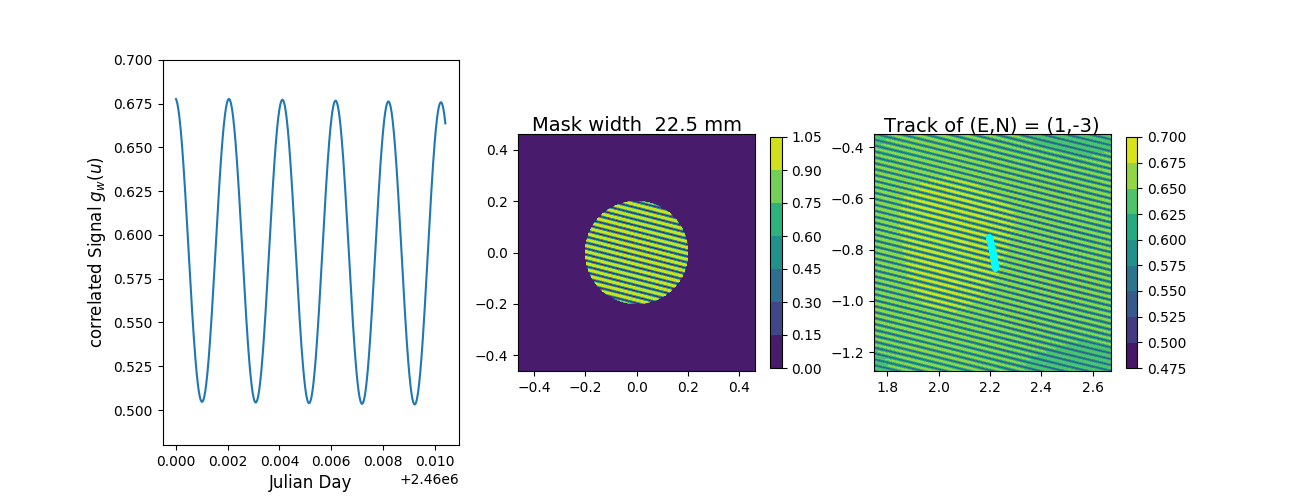}
		\caption{With value of mask's orientation $104.3^\circ$(1.82 radian).}
	\end{subfigure}
    \begin{subfigure}{\linewidth}
    	\includegraphics[width=\linewidth]{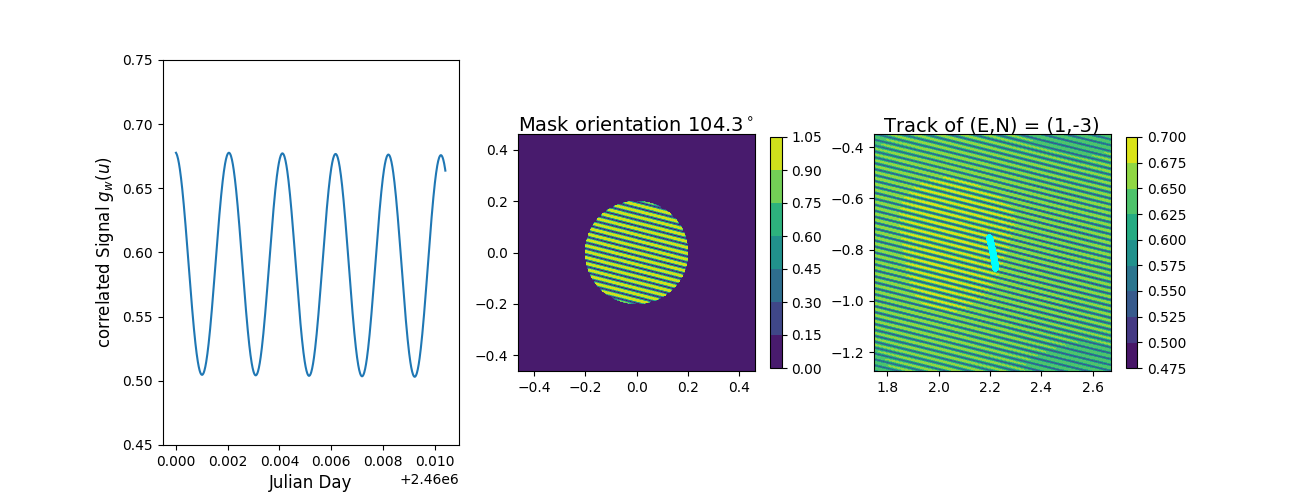}
    	\caption{With value of mask's width 22.5mm.}
    \end{subfigure}
    \caption{This figure shows the signal for the mask's strip width $b=22.5$mm (upper row) and for the mask's orientation $\vun = 104.3^\circ$ (lower row). The mask is shown in the middle panels. The underlying correlation $g(\vu)$ roughly corresponds to the position of \acen~AB in 2020, being observed from a latitude of $44^\circ\rm\,S$ at a wavelength of $\lambda=\SI{600}{\nano\metre}$.  The right panel shows $\gw(\vu)$ on the $\lambda (u,v)$ plane.  The cyan tracks correspond to a pair of telescopes with $(\Delta_{\rm E}=\SI{1}{\metre},\Delta_{\rm N}=\SI{-3}{\metre})$ over an interval of 15 minutes. The left panel shows $\gw(\vu)$ along the cyan tracks. The animated videos of variation in signal with $b$ and $\vun$ are available in the supplementary material.
    \label{fig:gif}}   
\end{figure*}

\begin{figure*}
	\centering
	\begin{subfigure}{0.50\linewidth}
		\includegraphics[width=\linewidth]{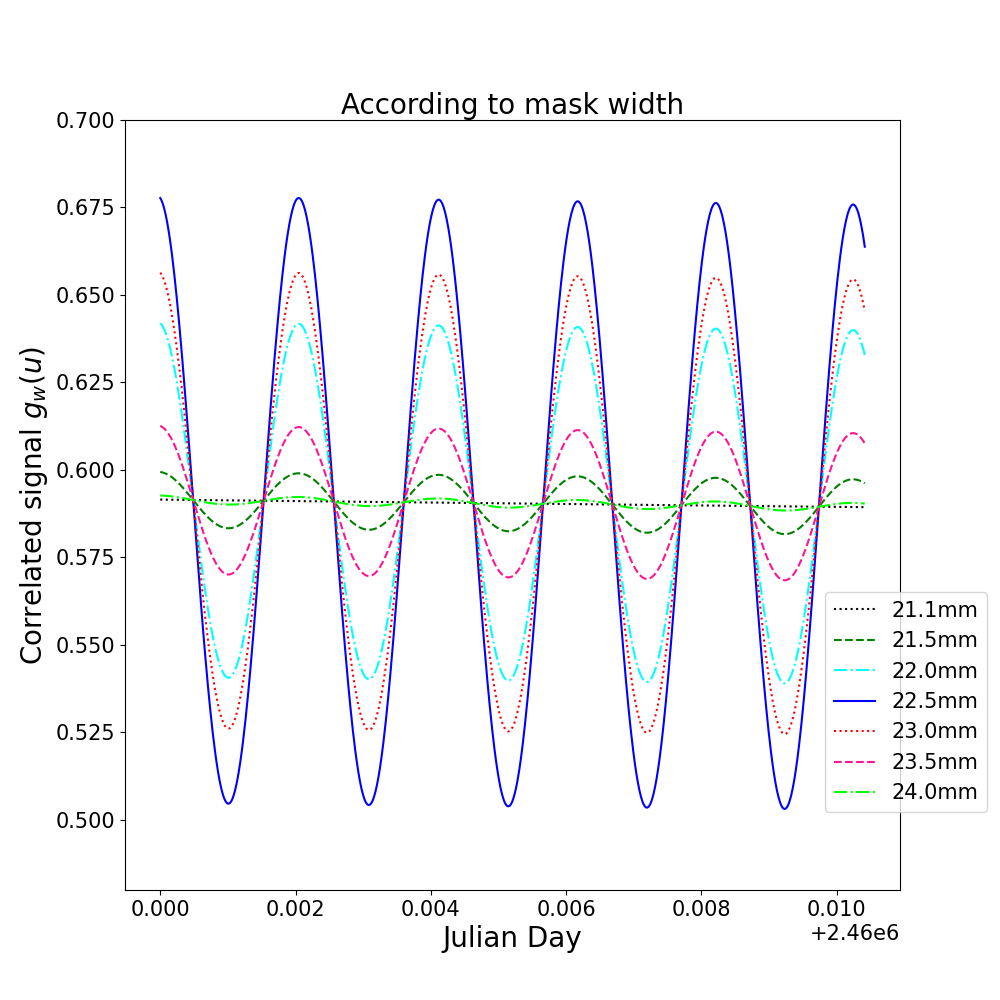}
		\caption{$\gw(\vu)$ for different values of $b$.}
		\label{fig:variwdth}
	\end{subfigure}\hfill
	\begin{subfigure}{0.50\linewidth}
		\includegraphics[width=\linewidth]{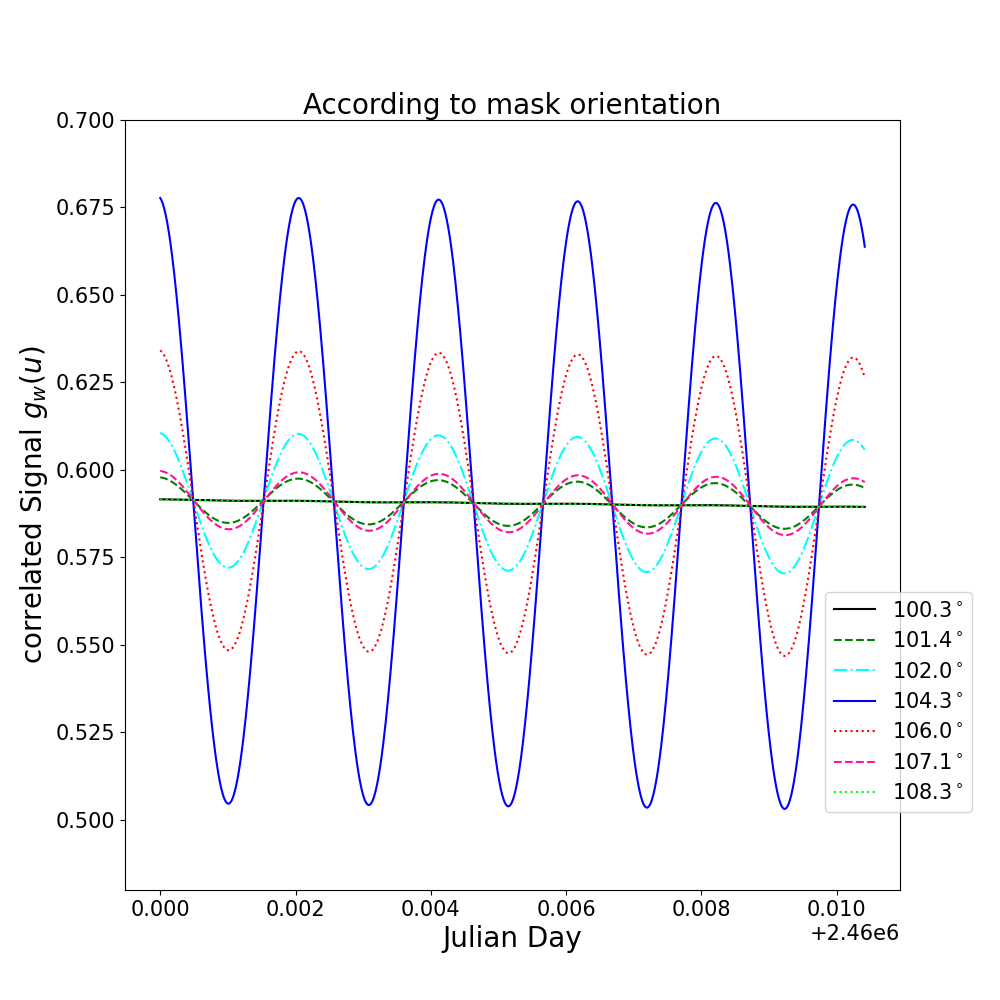}
		\caption{$\gw(\vu)$ for different values of $\vun$ .}
		\label{fig:variagl}
	\end{subfigure}\hfill
	\caption{This figure shows how the aperture mask affects the signal $\gw(\vu)$. In the left panel the stripe width $b$ is varied, and in the right panel, the mask orientation $\vun$ is varied. As in Fig~\ref{fig:gif}, the observation time is an interval of 15 minutes with baseline $(\Delta_{\rm E}=\SI{1}{\metre},\Delta_{\rm N}=\SI{-3}{\metre})$.\label{fig:variboth}}
\end{figure*}

\begin{figure*}
	\centering
	\begin{subfigure}{0.50\linewidth}
		\includegraphics[width=\linewidth]{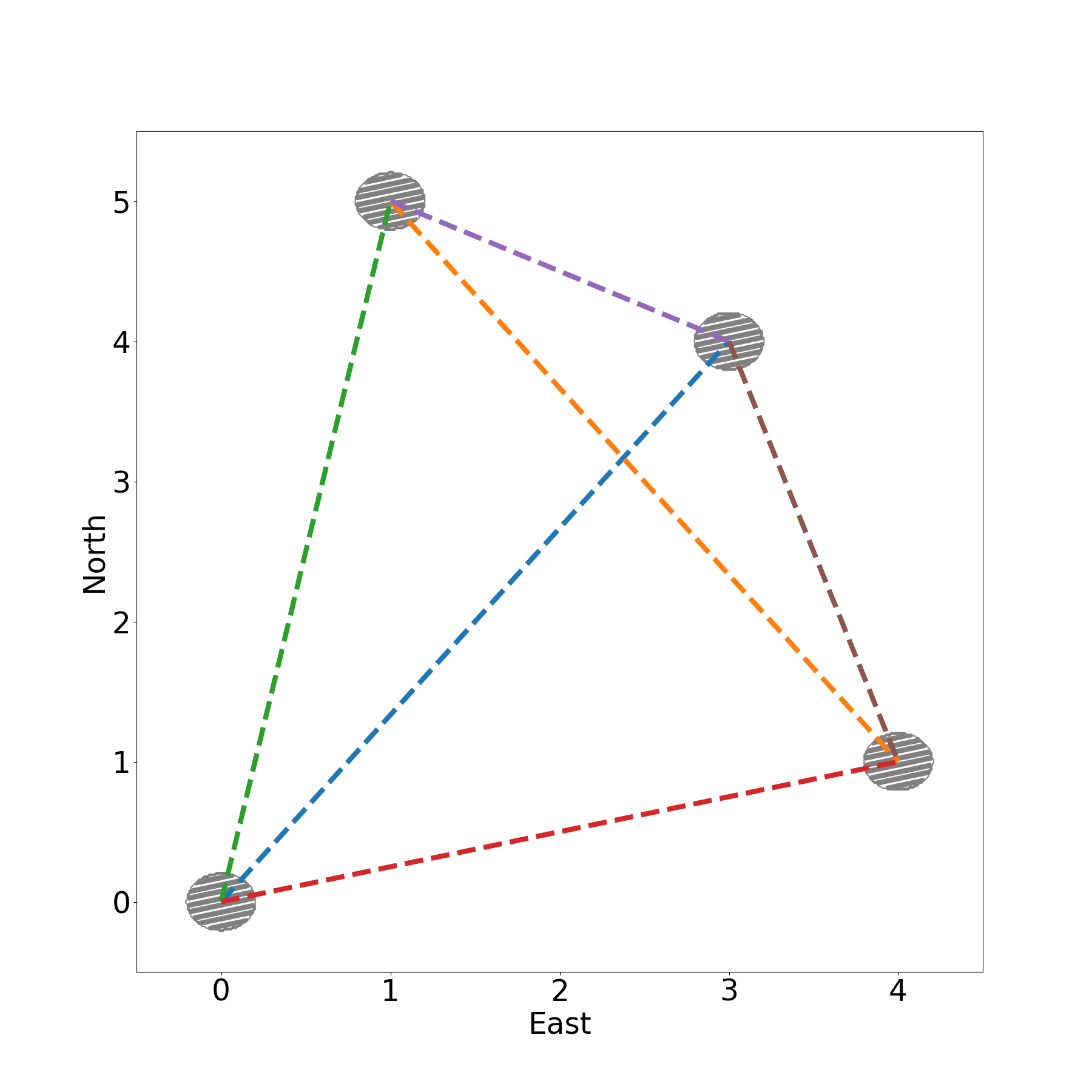}
		\caption{The position of telescopes}
		\label{fig:telscpos}
	\end{subfigure}\hfill
	\begin{subfigure}{0.50\linewidth}
		\includegraphics[width=\linewidth]{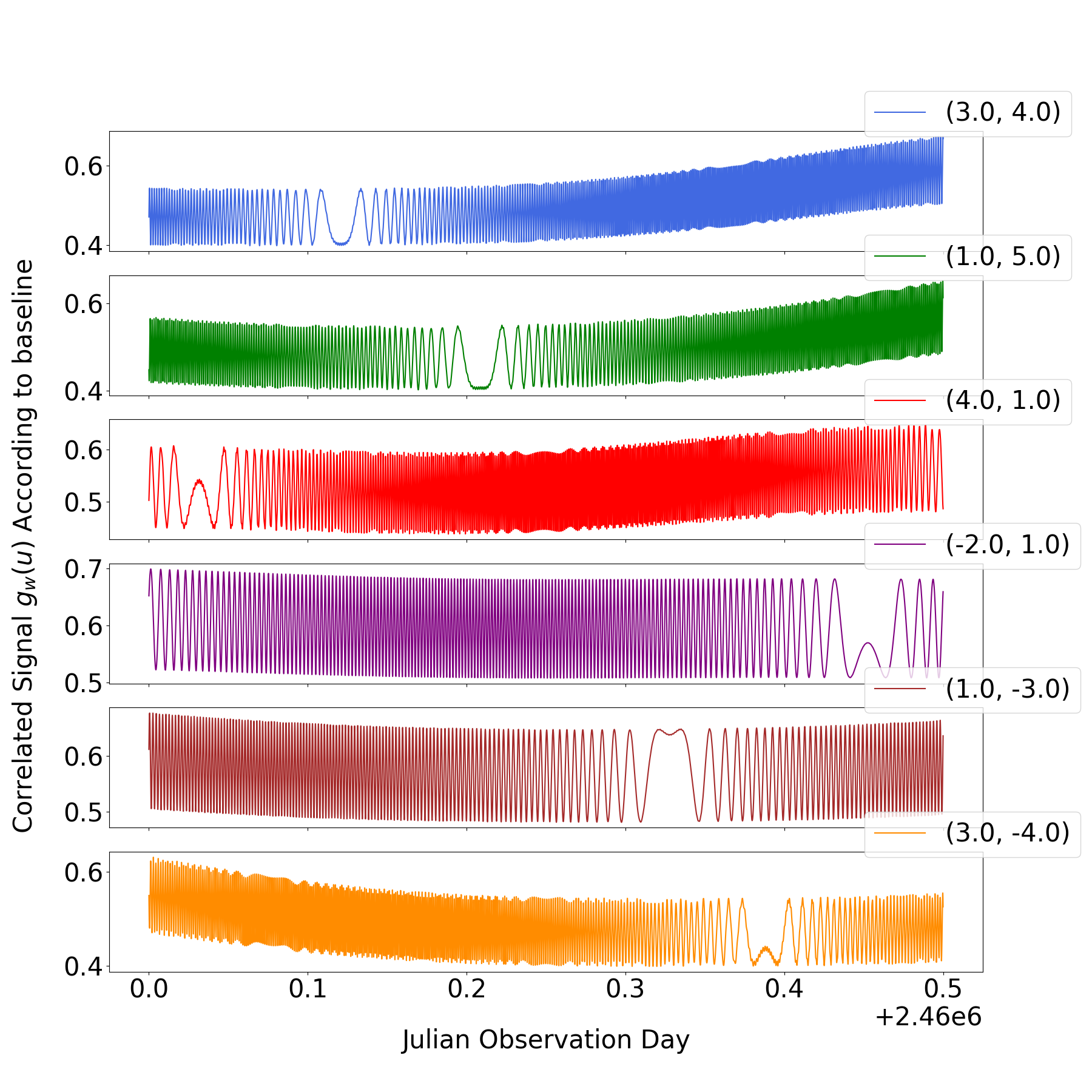}
		\caption{Model signal according to baselines}
		\label{fig:mdata}
	\end{subfigure}\hfill
	\caption{The left figure shows the position of the four telescopes ((E,N) = (0,0), (1,5), (3,4), (4,1)). The right figure shows the signal according to each baseline for one-night observation. Here the colors of baselines joining the telescope have been taken as the same colors as signals. }
\end{figure*}

Now, the objective plane of a telescope is obviously perpendicular to
the line of sight, and hence parallel to the $(u,v)$ plane.  We can
therefore express the aperture of a telescope as $W(\vu)$. The
finite-aperture signal is then $g(\vu)$ convolved with the apertures
of both telescopes. Thus we have
\begin{equation}
\gw(\vu) =
\frac{\int g(\vu+\vu_2-\vu_1) \,
           W(\vu_1) \, W(\vu_2) \; d^2\vu_1 \, d^2\vu_2}
     {\int W(\vu_1) \, W(\vu_2) \; d^2\vu_1 \, d^2\vu_2}
\label{eqn:correlated}
\end{equation}
for the finite aperture, and then
\begin{equation}
g^{\rm obs} = \frac{\Delta\tau}{\Delta t} \, \gw(\vu)
\label{eq:gmeas}
\end{equation}
will be the measured intensity correlation.

The convolution \eqref{eqn:correlated} suggests that making $W(\vu)$
similar to the fringe pattern itself will prevent the fringes from being
washed out.  Accordingly we choose
\begin{equation}
W(\vu) = 
\begin{cases}
  \cos^2 \Big((\pi\lambda/b) \, \vu\cdot\vun \Big)
  & \text{for}\ \rho\lambda < \tilde{a} \\
  0                & \text{otherwise}
\end{cases}
\label{eq:mask}
\end{equation}
which looks complicated, but really just describes a set of sinusoidal
stripes inside a circle.  Here $\tilde{a}$ denotes the radius of the telescope
aperture, $\vun$ is a unit vector and hence, specifies the orientation of the stripes, and $b$ is
the stripes width (the length of a stripe will vary according to its
position on aperture).

Computation of the convolution (\ref{eqn:correlated}) is done on a
$1024\times1024$ grid.

Figs.~\ref{fig:gif} and \ref{fig:variboth} illustrate the dependence
of the signal $\gw(\vu)$ on the stripe width $b$ and the orientation
$\vun$.  For the assumed relative position and the stars and an
observing wavelength of $\SI{600}{\nano\metre}$, the fringes are most
apparent in $\gw(\vu)$ for stripe width $b=\SI{2.25}{\centi\metre}$
and $\vun$ oriented at $104.3^{\circ}$ with respect to $\vu$. The
signal degrades noticeably when $b$ and $\vun$ are even a few percent
from the optimal values.  Hence, advance knowledge of the current
relative positions of the stars to $\sim1\%$ is desired, in order to
optimise the aperture mask. This is unlikely to be a problem in
practice for a wide binary such as the \acen~AB.

A possible additional complication, depending on the type of
telescope mounting, is that the mask may need to be rotated as the
telescope tracks the target across the sky, to compensate for field
rotation.  Equatorial mounting would eliminate this issue.

Four telescopes have been considered during the simulation work.
Fig.~\ref{fig:telscpos} shows their arrangement. Each telescope has a
diameter of $\SI{40}{\centi\metre}$ and a mask with the optimal
aperture mask mentioned earlier. The signal for one-night of observation from all
six baselines formed by these four telescopes is shown in
Fig.~\ref{fig:mdata}.

\subsection{Signal to Noise}

In a resolution time $\Delta t$ the number of HBT correlated photon
pairs will be
\begin{equation}
S = (A \Phi)^2 \; \gw(\vu) \times \frac{\Delta t}{\Delta\tau}
\label{eqn:signal}
\end{equation}
where
\begin{equation}
A = \int W(\vu) \; d^2\vu
\end{equation}
is the effective area of each telescope, $\Phi$ is the spectral photon
density, and $\Delta\tau$ is the coherence time.

The coherence time is the reciprocal of the frequency bandwidth, thus
$1/\Delta\nu=\lambda^2/(c\Delta\lambda)$.  For the assumed observing
wavelength of $\SI{600}{\nano\metre}$ the wave period is
$\lambda/c=\SI{2.0}{\femto\second}$.  No specific $\Delta\lambda$ is
assumed, but narrow bandwidths are generally desirable.  A bandwidth
of $\Delta\lambda=\SI{1}{\nano\metre}$ gives
$\Delta\tau=\SI{1.2}{\pico\second}$.  In practice, $\Delta t$ will be
orders of magnitude larger than $\Delta\tau$.  The photon spectral
density denotes the photons per unit area in a coherence time
$\Delta\tau$ in one polarisation, which for a thermal source is given
by
\begin{equation}
  \Phi = \frac{(\nu/c)^2}{e^{z}-1} \times \Delta\Omega
  \qquad z = \frac{h\nu}{kT}
\end{equation}
where $\Delta\Omega$ is the solid-angle area of the source, and $T$ is
the effective temperature.  The spectral energy flux (measured in Jy) is
$2h\nu\Phi$.

In an interval $\Delta t$, the noise due to chance coincidences will
be
\begin{equation}
N = A \Phi \times \frac{\Delta t}{\Delta\tau}\,.
\label{eqn:noise}
\end{equation}
Dividing equations~\eqref{eqn:signal} and \eqref{eqn:noise} gives the
signal to noise of
\begin{equation}
{\rm SNR}_{\Delta t} = A \Phi \; \gw(\vu)
\end{equation}
in one time slice $\Delta t$.   We assume $\Delta t=\SI{100}{\pico\second}$.  

In practice, the correlation would be done in hardware and averaged
over an $\tav$ during which $\vu$ can be assumed constant. We
take $\tav=\SI{1}{\second}$. This constitutes a data point for
fitting. The SNR of such a data point will be
\begin{equation}
{\rm SNR}_{\tav} = A \Phi \; \gw(\vu) \; (\tav/\Delta t)^{1/2}\,.
\label{eqn:truesig}
\end{equation}
For simulations, one can take the signal to be $\gw$ and the noise to be
\begin{equation}
N = (\Delta t/\tav)^{1/2} (A \Phi)^{-1} \,.
\label{eqn:truenoise}
\end{equation}

It is noted that $\gw(\vu)$ is several times smaller than $g(\vu)$.
That is, the aperture mask sacrifices a lot of signal. Let us
therefore consider what the SNR would be, if small light buckets
without aperture masking were used. For a given effective
temperature, $\Phi\propto (R/D)^2$, where $R$ is the radius of the
larger star and $D$ the distance.  If $a$ is the separation between
the stars, the width of the interference fringes will be $\propto D/a$.
Without aperture masking, the mirror diameter must to kept smaller than
the fringe width, implying $A\propto (D/a)^2$.  This would give
${\rm SNR} \propto (R/a)^2$, which becomes hopelessly small for wide binaries.
Pointing error is not considered.

\section{Orbits}\label{sec:orbits}

If the \acen\ system has a planet, the observed astrometry of the two
bright stars would be part of a four-body system, including the planet
and Proxima. We have accordingly carried out orbit integrations of a
four-body system of the three stars and a fictitious planet around
\acen~A, with many choices for the planet mass and orbit.  Further
perturbations, such as the Galactic tidal field, are not expected to
be significant, and are not included.  The initial conditions are
based on orbital parameters from the literature for \acen~A and B
\citep{2002A&A...386..280P}, and for Proxima with respect to the
barycentre of A and B \citep{2017A&A...598L...7K}.

\begin{table*}
\centering
\begin{tabular}{cccccccc}
\hline\hline
$M$ ($M_\odot$) & $P$ (yr) & $a$ (au) & $e$ & $\iota$ (deg) & $\Omega$ (deg) & $t_{\rm peri}$ (yr) & $\omega$ (deg) \\
\hline
1.1 \\
$10^{-2}$ &    1    &    1    & 0      & 80  &   0    &    0    &   0     \\
0.9       &   79.9  &   23.5  & 0.518  & 79  & 204.85 & 1875.66 & 231.65  \\
0.1221    & 547000  &  8700   & 0.5    & 107 & 126    &  285000 &  72     \\
\hline\hline
\end{tabular}
\caption{Masses $M$ and orbital elements (Period $P$, semi-major axis $a$, eccentricity $e$, inclination $\iota$, longitude of the ascending node $\Omega$, periastron epoch ($t_{\rm peri}$), and argument of periapsis $\omega$) for the \acen\ system.  The top
  line is for \acen~A.  The second line represents a fictitious
  planet.  The third line gives the observed orbit of the \acen~AB
  system \citep{2002A&A...386..280P}, while the last line gives the
  orbit of Proxima Centauri with respect to the barycentre of the two bright
  stars \citep{2017A&A...598L...7K}. \label{tab:orbitelems}}
\end{table*}

\subsection{The Orbital Integration}

The equations of motion of an $N$-body system with masses $m_i$
($i=0,1,\ldots,N-1$) in terms of the positions $\ri$ and velocities $\vi$
in an inertial reference frame are given by:
\begin{equation}
\ddot{\ri} \equiv \dot{\vi} = - \sum_{j=0; j \neq i}^{N-1} Gm_j
\frac{\ri- \rj}{|\ri-\rj|\cubed} \label{eqn:EqnOfMotion}
\end{equation}
where $G$ is the gravitational constant.

We use the well known Leapfrog algorithm, also otherwise known as
Drift-Kick-Drift, or the Verlet method, to solve for the orbits of
this system. Under this algorithm, the set of phase space vectors
$\{(\ri(t), \vi(t)), i=0,1,\ldots, N-1\}$ are advanced by the following set
of steps:
\begin{equation}
\begin{aligned}
\ri^{\rm int} &=  \ri^{\rm old} + \ \vi^{\rm old} \times \half t_{\rm step} \\
\vi^{\rm new} &= \vi^{\rm old} - {\sum_{j=0,j \neq i}^{N-1}  Gm_j}
\frac{\ri^{\rm int} - \rj^{\rm int}}{\vert \ri^{\rm int}- \rj^{\rm int} \vert\cubed} \,
\times t_{\rm step}\\
\ri^{\rm new} &=  \ri^{\rm int} +  \ \vi^{\rm new} \times \half t_{\rm step}.
\end{aligned}
\label{eqn:DKD}
\end{equation}
The first step (Drift) in equation~\eqref{eqn:DKD} produces an interim position vector $\ri^{\rm int}$ using the present values of the position vector $\ri^{\rm old}$ and the velocity vector $\vi^{\rm old}$ over half the chosen time step, {\it i.e.,} $\frac{1}{2} t_{\rm step}$. The second step (Kick) is produced by generating the next set of velocity vectors $\vi^{\rm new}$ over a full time step $t_{\rm step}$ by using the accelerations evaluated at the interim position vectors $\ri^{\rm int}$. Finally, in the last step (a second Drift), the next set of position vectors $\ri^{\rm new}$ are evaluated with the remaining half a time step $\frac{1}{2} t_{\rm step}$ using the set of velocity vectors $\vi^{\rm new}$ at the next time step.
More efficient algorithms are available \citep{2020gfbd.book.....M}, but Leapfrog is sufficient for our purposes.  It is known to conserve momentum and angular momentum exactly.  The total energy has an oscillatory variation proportional to $(t_{\rm step})^2$ and no secular variation.

Light-second units have been used for the computation.  That is done
by making the code variables $\ri/c$ and $\vi/c$, and mass parameters in
the program $Gm_i/c^3$.  The time step is $t_{\rm step}=\SI{1e6}{\second}$.

\subsection{Initial conditions}

Table~\ref{tab:orbitelems} gives the masses and orbital elements of
the stars, taken from the literature, and (in the second) line the
parameters for a fictitious planet.

Orbital elements such as in Table~\ref{tab:orbitelems} describe
instantaneous two-body orbits.  They can be transformed to cartesian
positions and velocities by the following procedure \citep[for a
  derivation, see e.g.,][]{SahaTaylor}.

First we solve Kepler's equation
\begin{equation}
  \psi - e\sin\psi = \frac{2\pi}P (t-t_{\rm peri})
\label{eqn:kepler}
\end{equation}
for the eccentric anomaly $\psi$ with $t_{\rm peri}$ being the epoch of periastron of the system.  We then use the value of $\psi$ to
compute the position in the orbital plane, and then we apply
three rotations
\begin{equation}
  \begin{pmatrix}
     x \\ y \\ z
  \end{pmatrix}
  =
  a \times R_3(\Omega) \, R_1(\iota) \, R_3(\omega)
  \begin{pmatrix}
     \cos\psi-e \\ \sqrt{1-e^2}\sin\psi \\ 0
  \end{pmatrix}
  \label{eqn:rotate}
\end{equation}
so as to orient the coordinates conveniently with $z$ along the line
of sight. In equation~\eqref{eqn:rotate}, the set of parameters \{$a$, $e$, $\iota$, $\Omega$ and $\omega$\} represent the standard Keplerian orbit elements noted in table~\ref{tab:orbitelems}. The $R_1$ matrices are as in equation~\eqref{eqn:R1}, and $R_3$ is
given by cyclically permuting the indices.  The analogous expression
for velocities follows by noting
\begin{equation}
  \frac{d\vec{x}}{dt} = \frac{2\pi}P \frac{d\vec{x}}{d\psi}
\end{equation}
which itself follows from equation~\eqref{eqn:kepler}.

Applying the above procedure to the orbital elements, gives us three
positions $\vec{x}_1,\vec{x}_2,\vec{x}_3$ and three corresponding
velocities.  These are two-body coordinates and velocities, and
correspond to pairs of bodies, not individual bodies.  We now
interpret these as Jacobi variables.  That is, $\vec{x}_1$ is
considered as the position of the second body (the fictitious planet)
with respect to the first body (\acen~A), $\vec{x}_2$ is taken from
the third body (\acen~B) to the barycentre of the earlier bodies, and
so on.  Inertial coordinates $\ri^I$ are then given by
\begin{equation}
\begin{aligned}
  \vec{X}_0^I &= 0 \\
  \ri^I &= \vec{x}_i + \frac{\sum_{j<i} m_j \rj^I}{\sum_{j<i} m_j}
\end{aligned}
\end{equation}
and similarly for velocities.  Finally the positions and velocities
are transformed to barycentric coordinates
\begin{equation}
  \ri = \ri^I - \frac{\sum_j m_j \rj^I}{\sum_j m_j}
\end{equation}
and again similarly for velocities.

\subsection{The simulated orbits}

\begin{figure*}
	\centering
	\begin{subfigure}{0.50\linewidth}
		\includegraphics[width=\linewidth]{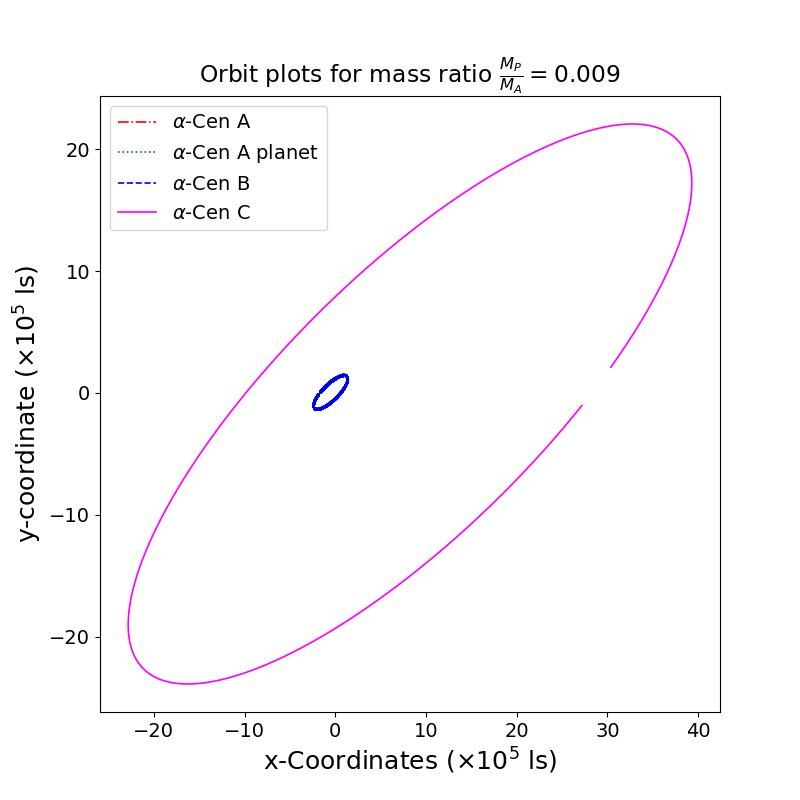}
		\caption{Orbit of \acen~system}
		\label{fig:Binary plot1}
	\end{subfigure}\hfill
	\begin{subfigure}{0.50\linewidth}
		\includegraphics[width=\linewidth]{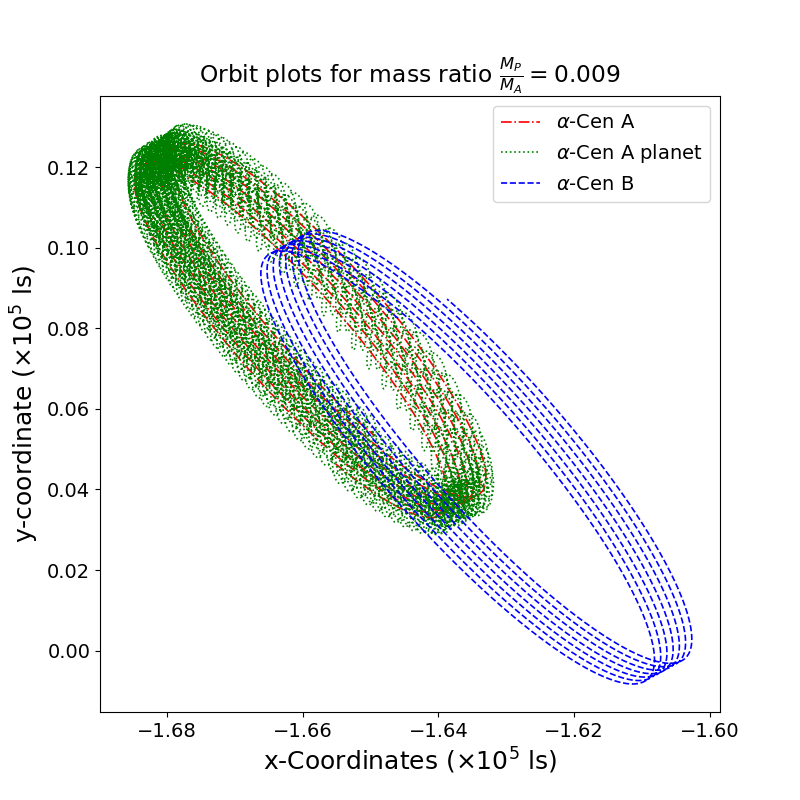}
		\caption{Orbit of \acen~AB system with a planet}
		\label{fig:Binary plot2}
	\end{subfigure}\hfill
	\caption{Left: a four-body system (\acen~system with a planet) for the time duration $\approx$ 0.545 Myr. Right: the orbit of the binary system ($\alpha$-Cen AB) with the planet for a period of $\approx$ 580 yr. The parameters of these objects are provided in Table~\ref{tab:orbitelems}.}
	\label{fig:binary_plot}
\end{figure*}

\begin{figure*}
	\centering
	\begin{subfigure}{0.50\linewidth}
		\includegraphics[width=\linewidth]{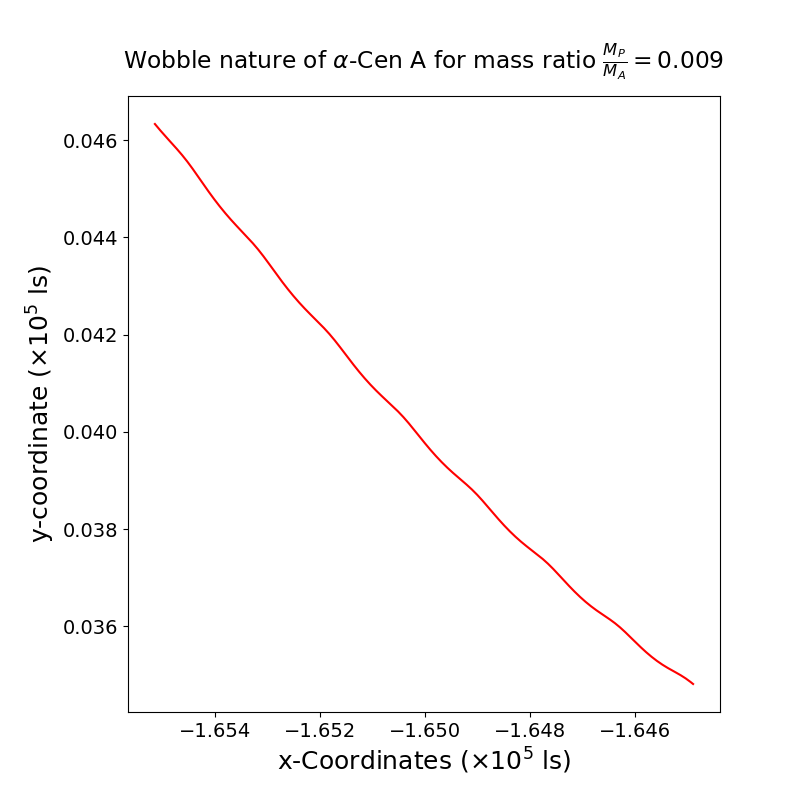}
		\caption{Orbit of \acen~A for $\approx$ 6 yr of orbit.}
		\label{fig:alphaA}
	\end{subfigure}\hfill
	\begin{subfigure}{0.50\linewidth}
		\includegraphics[width=\linewidth]{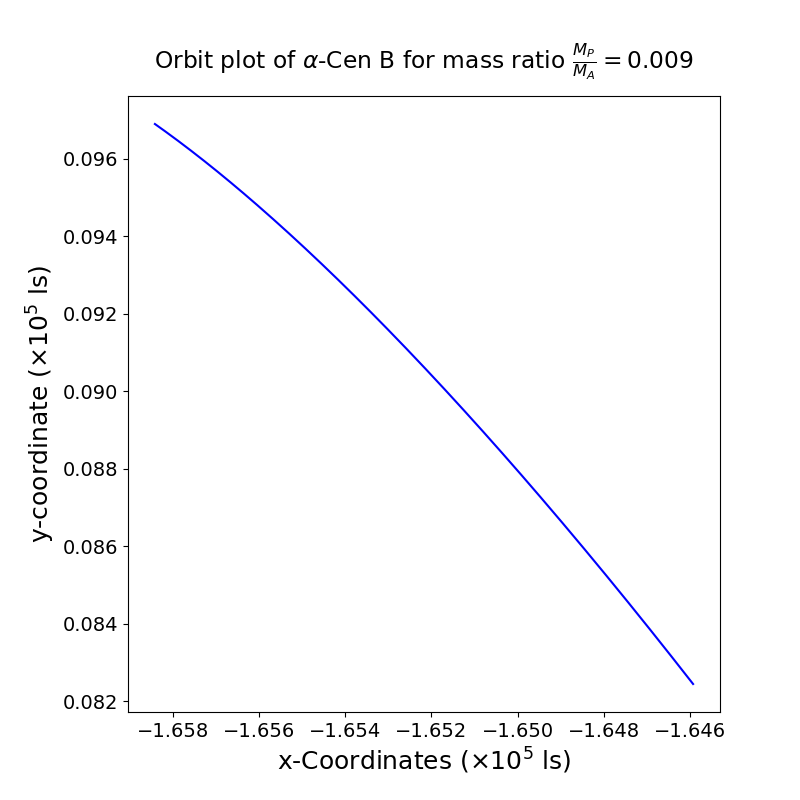}
		\caption{Orbit of \acen~B for $\approx$ 6 yr of orbit.}
		\label{fig:alphaB}
	\end{subfigure}\hfill
	\caption{This figure is the orbit plot of \acen~A and \acen~B for $\approx$ 6 yr of orbit. The left figure shows the perturbation in the orbit of \acen~A as the ten Jupiter mass planet is revolving in 1 AU semi-major axis orbit around it.  The perturbation in the trajectory of \acen~B (right figure) is not perceptible as (1) there is no planet around it and (2) the perturbations due to the perturbed motion of \acen~A is expected to be of higher order.}
	\label{fig:perturbation_plot}
\end{figure*}

Fig~\ref{fig:binary_plot} and Fig~\ref{fig:perturbation_plot} show the result of integration of equation~\eqref{eqn:EqnOfMotion} using leapfrog method. The initial condition and other parameters of all four objects have been taken from table~\ref{tab:orbitelems}. Fig~\ref{fig:binary_plot} shows the orbit of \acen~A, \acen~B, and planet (10 Jupiter mass) in red (dashed-dotted), blue (dashed), and green (dotted) colors, respectively. There is also an orbit of Proxima Centauri in magenta color (solid line), which is situated very far away from the barycenter of these three objects. These orbital solution is obtained by integrating the equations of motion equation~\eqref{eqn:DKD} for a period of $\approx$ 0.545 Myr. The orbit of the \acen~AB system is not distinguishable in Fig~\ref{fig:Binary plot1} as the separation of Proxima Centauri from the binary system is large enough. The orbit plot of \acen~AB with a planet has been shown in Fig~\ref{fig:Binary plot2}. As the Proxima Centauri is gravitationally bounded with \acen~AB binary \citep{2017A&A...598L...7K}, the orbit's shift of the binary system comes into the picture as a result. 

Fig~\ref{fig:alphaA} shows the trajectory of \acen~A with respect to the centre of mass of the \acen~AB -plus-planet system for a period of $\approx$ 6 years. The planet (with ten Jupiter mass) revolves around \acen~A in 1 AU semi-major axis; the perturbation effect can be seen in this figure. Fig~\ref{fig:alphaB} shows the trajectory of \acen~B for the same period. It may be expected that the perturbation in the trajectory of \acen~B may be imperceptible as (1) there is no planet around it and (2) the perturbations due to the perturbed motion of \acen~A should be of higher order.

\section{Results}\label{sec:Results}

\begin{figure}
\centering
\includegraphics[width=\linewidth]{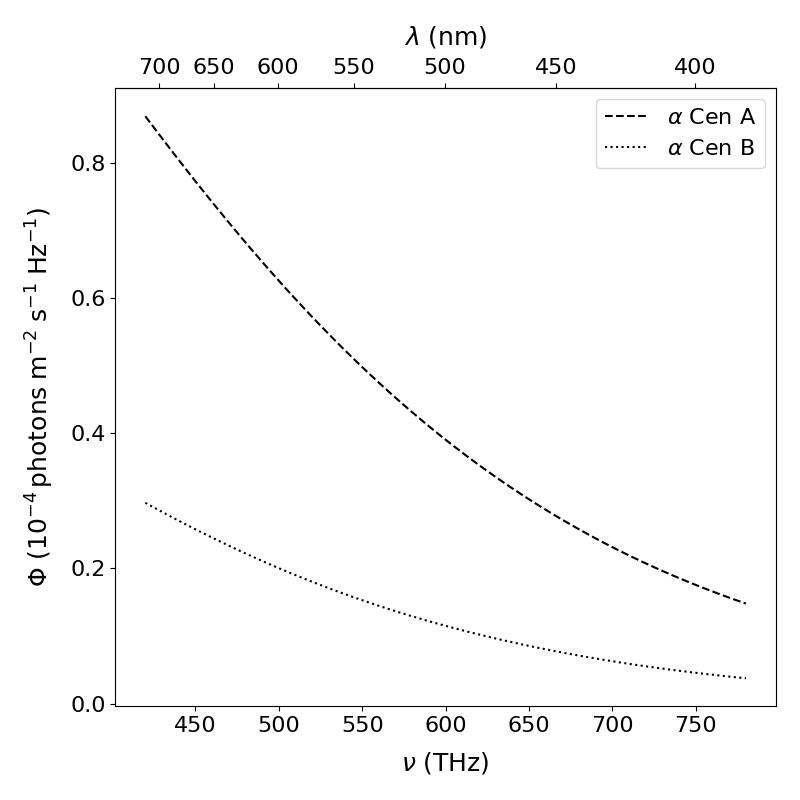}
\caption{Spectral photon flux in each polarisation for two blackbody
  sources approximating \acen~A and B.  These are simply the spectral
  energy density divided by $2h\nu$.  Both curves peak in the infrared
  range, outside the range of the figure. The quantity $\Phi$ used in
  the simulations is the sum of the two sources evaluated at
  $\SI{600}{\nano\metre}$.}
\label{fig:phspec}
\end{figure}

\begin{figure}
	\centering
	\includegraphics[width=\linewidth]{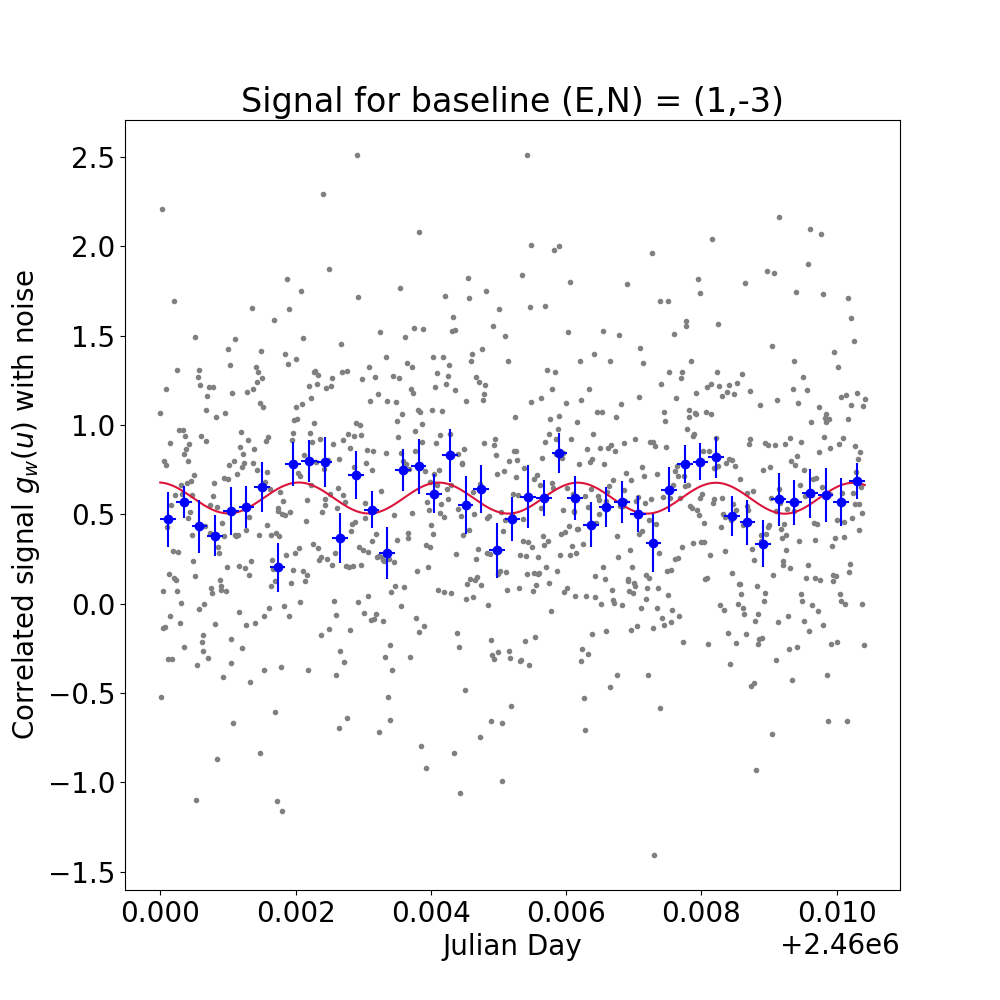}
	\caption{This figure shows the signal $\gw(\vu)$ and noise for
          one baseline with 10 spectral channels. The duration shown
          is $15\,$min. The grey dots are noisy data in bins of
          $1\,$s.  The blue error bars represent the noisy data in
          bins of $20\,$s, and the crimson line represents the
          noiseless signal.}
	\label{fig:simudata}
\end{figure}

\begin{figure}
\centering
\includegraphics[width=\linewidth]{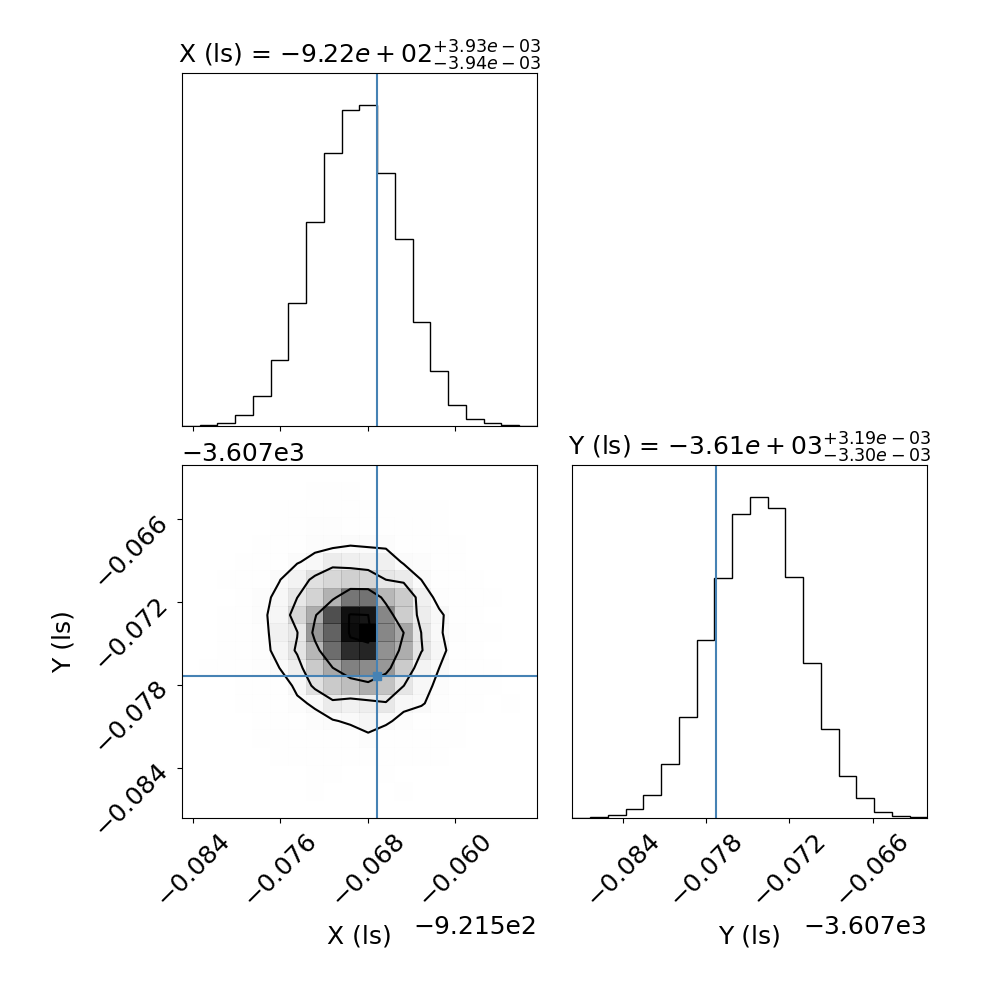}
\caption{Estimation of the sky-projected relative position $(X,Y)$
  \acen~A,B, with SNR corresponding to observing with 10 spectral
  channels for 1000 nights.  The error is less than $10\,\mu\rm as$
  (which corresponds to $\SI{7e-3}{\rm ls}$ at the distance of \acen).
\label{fig:estimation}}
\end{figure}

\begin{figure*}
	\centering
	\begin{subfigure}{0.50\linewidth}
		\includegraphics[width=\linewidth]{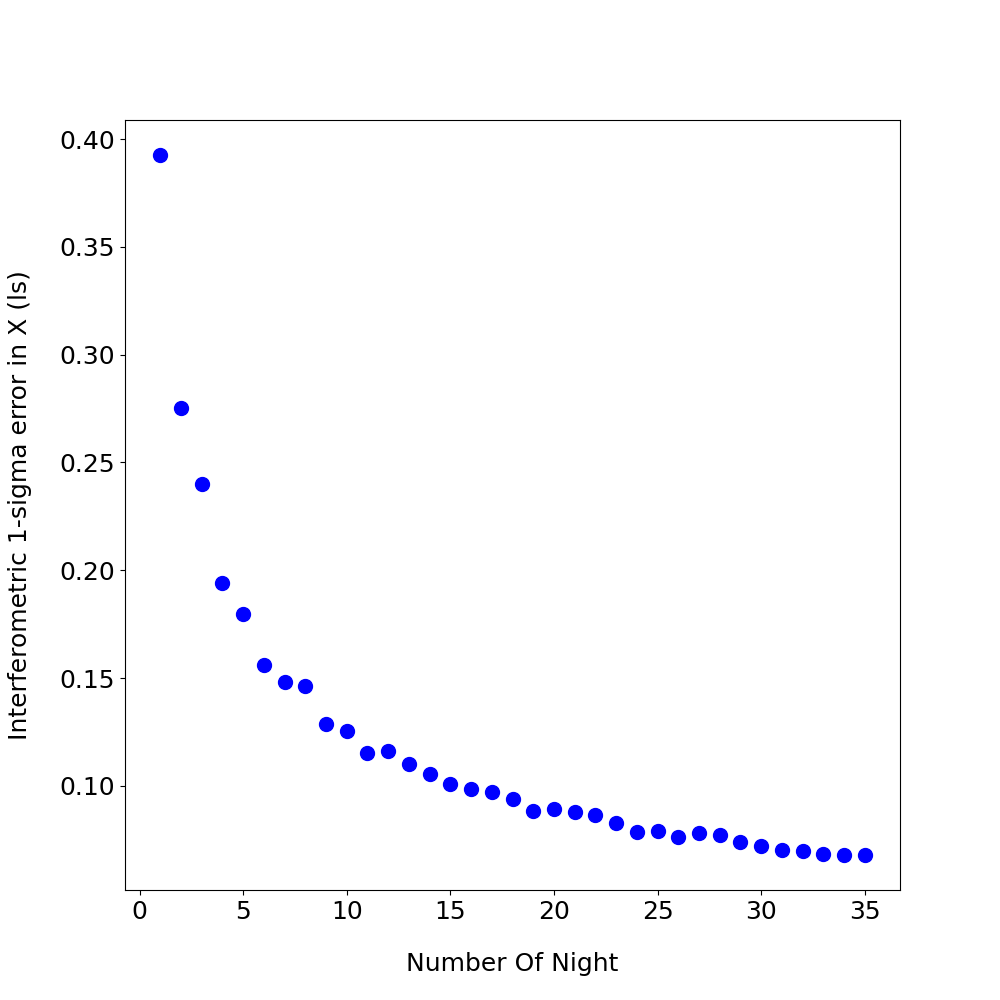}
		\label{fig:Xerror}
	\end{subfigure}\hfill
	\begin{subfigure}{0.50\linewidth}
		\includegraphics[width=\linewidth]{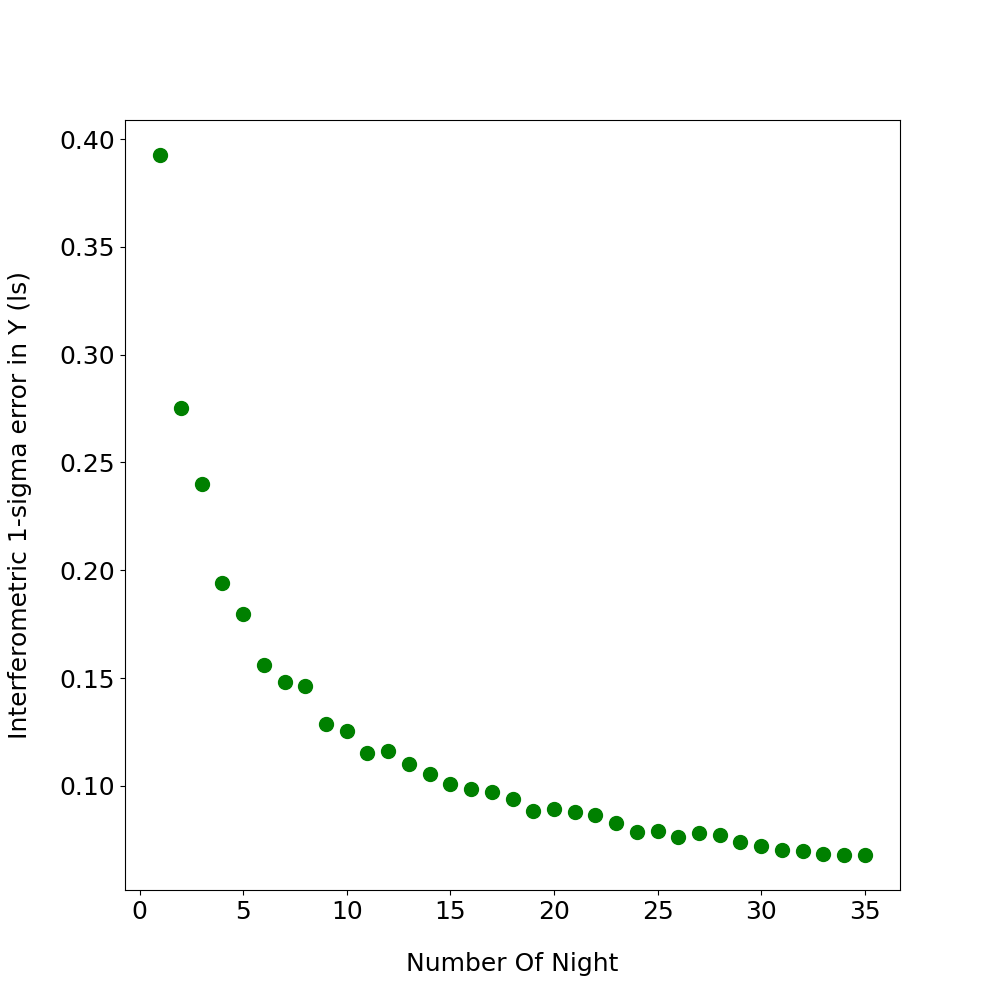}
		\label{fig:Yerror}
	\end{subfigure}\hfill
\caption{This figure shows the reduction in astrometric error with an increasing number of nights of observation.  One channel is assumed.  The left panel shows the error in the $X$ coordinate of the binary, and the right panel shows the error in $Y$.}
	\label{fig:error}
\end{figure*}

\begin{figure}
	\includegraphics[width=\linewidth]{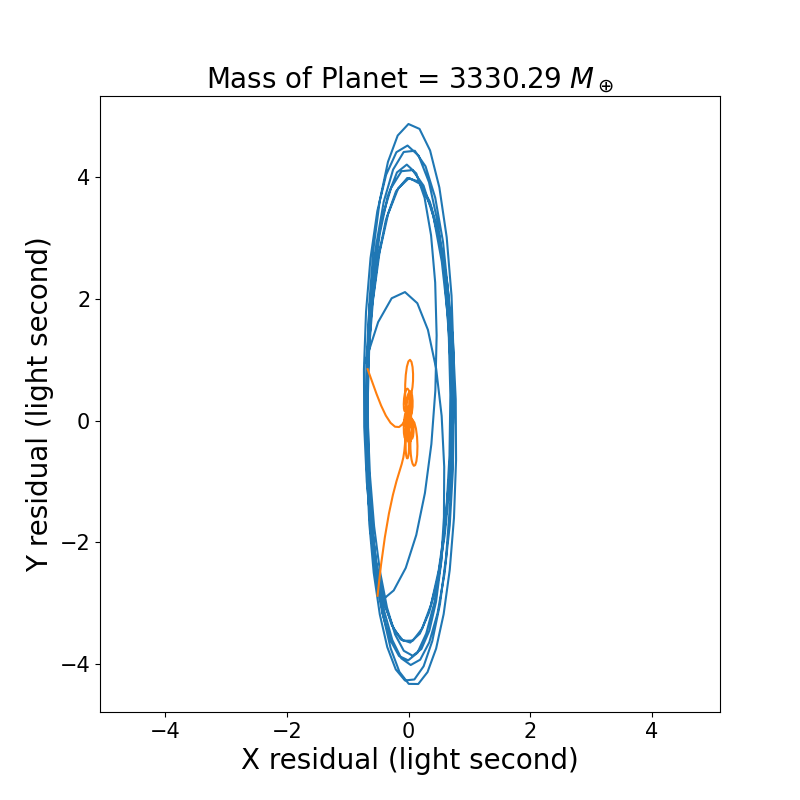}
	\caption{Astrometric wobble over approximately $6\,$yr,
          corresponding to the orbit shown in Fig.~\ref{fig:alphaA}.
          Blue:~residuals ($\Xwob(t),\Ywob(t)$) after removal of a
          polynomial up to $t^6$. Orange:~residuals
          ($\Delta\Xwob(t),\Delta\Ywob(t)$) after further removal of the
          largest periodic part. This example assumed an implausibly
          high planet mass ($\approx 10\,M_{\rm Jup}$) to make the wobble  
          perceptible (Fig.~\ref{fig:alphaA}).}
	\label{fig:leastsquare}
\end{figure}

\begin{figure*}
	\centering
	\begin{subfigure}{0.50\linewidth}
		\includegraphics[width=\linewidth]{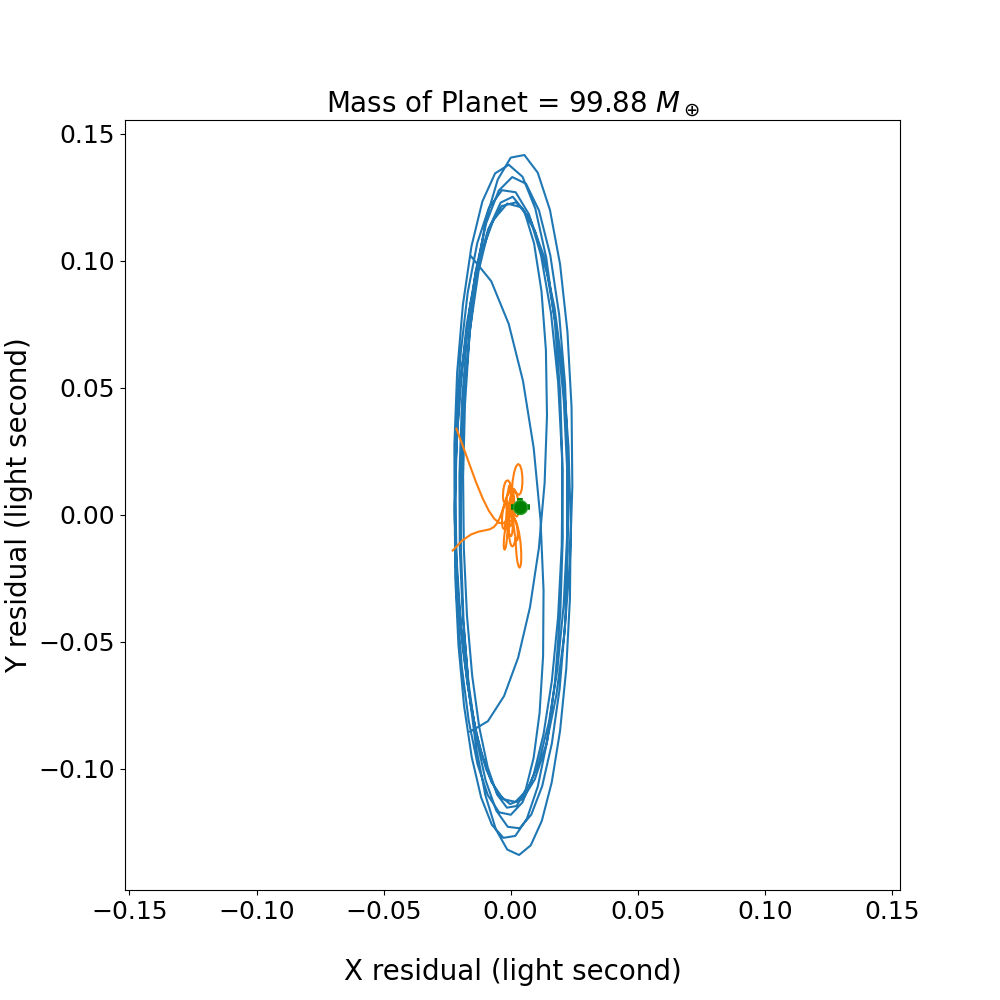}
		\caption{For a Saturn mass planet}
		\label{fig:mass1}
	\end{subfigure}\hfill
	\begin{subfigure}{0.50\linewidth}
		\includegraphics[width=\linewidth]{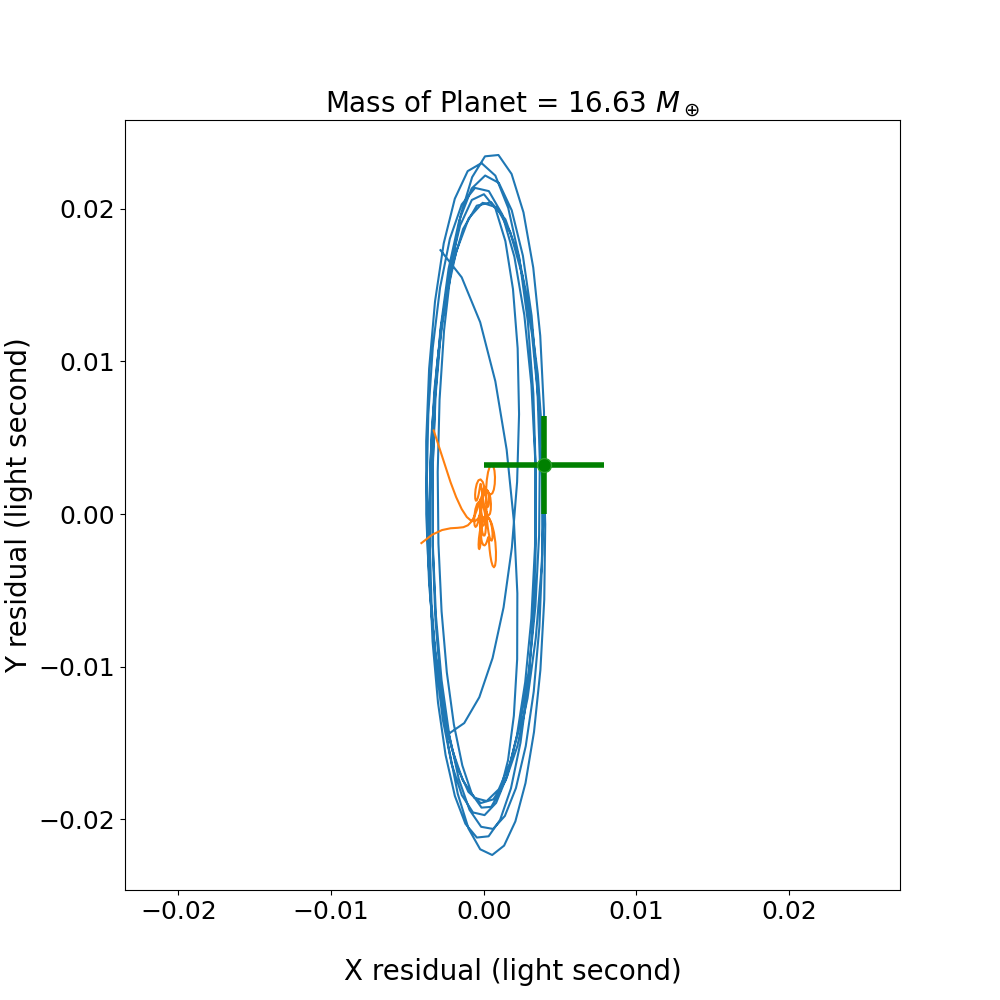}
		\caption{For a Neptune mass planet}
		\label{fig:mass2}
	\end{subfigure}\hfill
    \caption{This figure shows the residuals with an interferometric error bar. The interferometric bar and also residuals have been taken for different masses.The animated video of residuals with different planet masses is available in the supplementary material.}
	\label{fig:massgif}
\end{figure*}

\begin{figure*}
	\centering
	\begin{subfigure}{0.50\linewidth}
		\includegraphics[width=\linewidth]{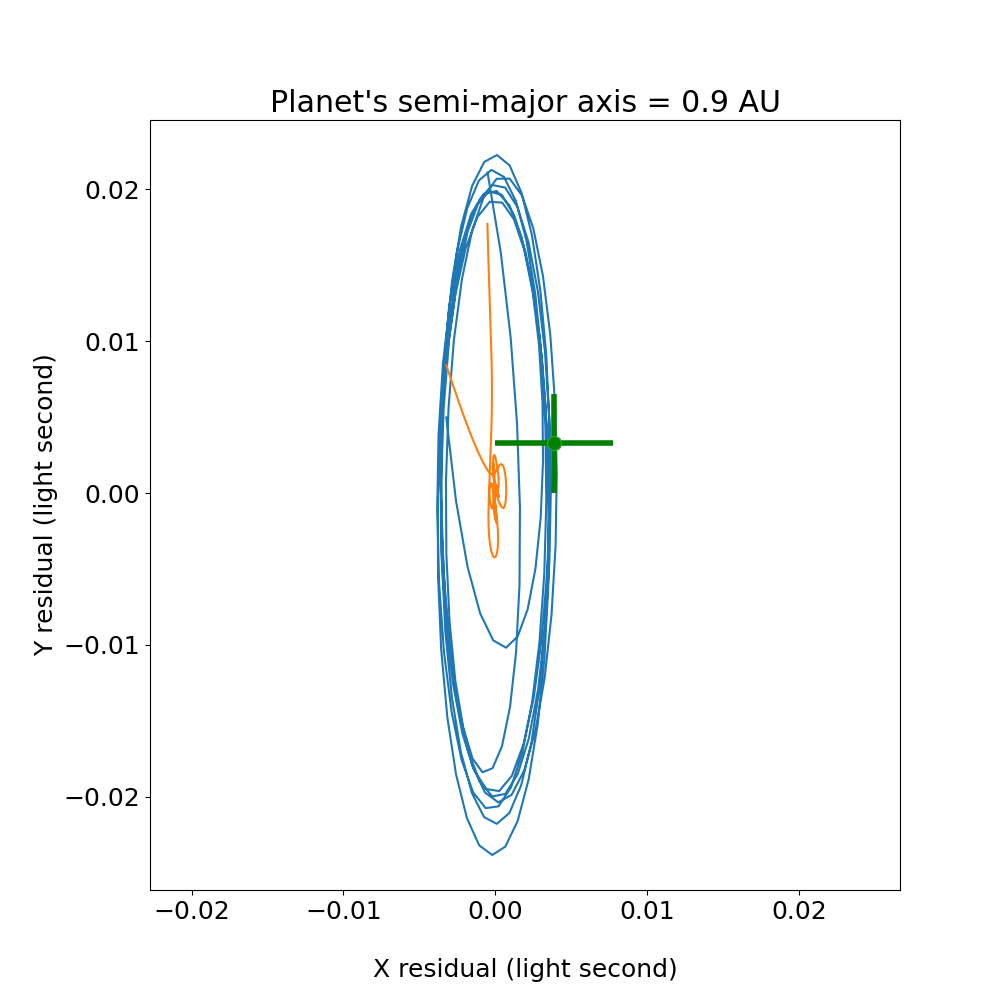}
		\caption{Planet's estimated period = 0.841 yr}
		\label{fig:AU1}
	\end{subfigure}\hfill
	\begin{subfigure}{0.50\linewidth}
		\includegraphics[width=\linewidth]{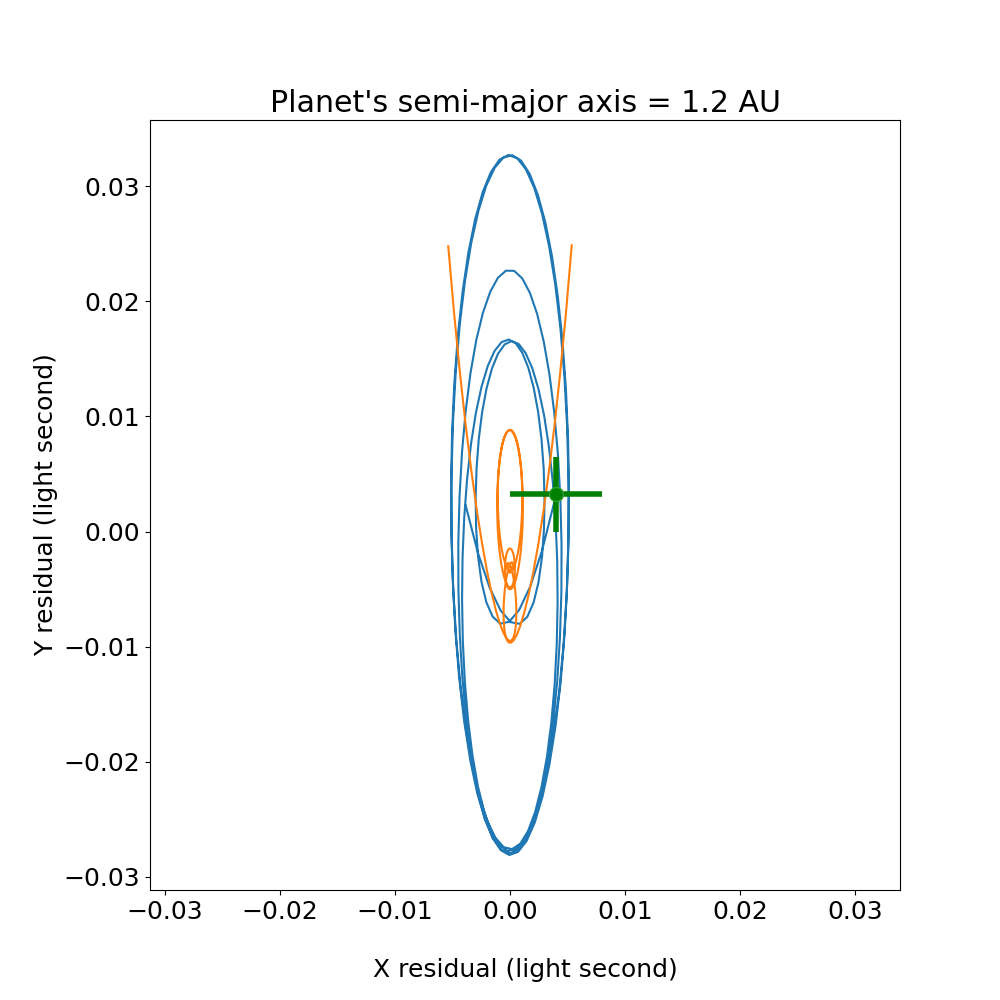}
		\caption{Planet's estimated period = 1.281 yr}
		\label{fig:AU4}
	\end{subfigure}\hfill
	\begin{subfigure}{0.50\linewidth}
		\includegraphics[width=\linewidth]{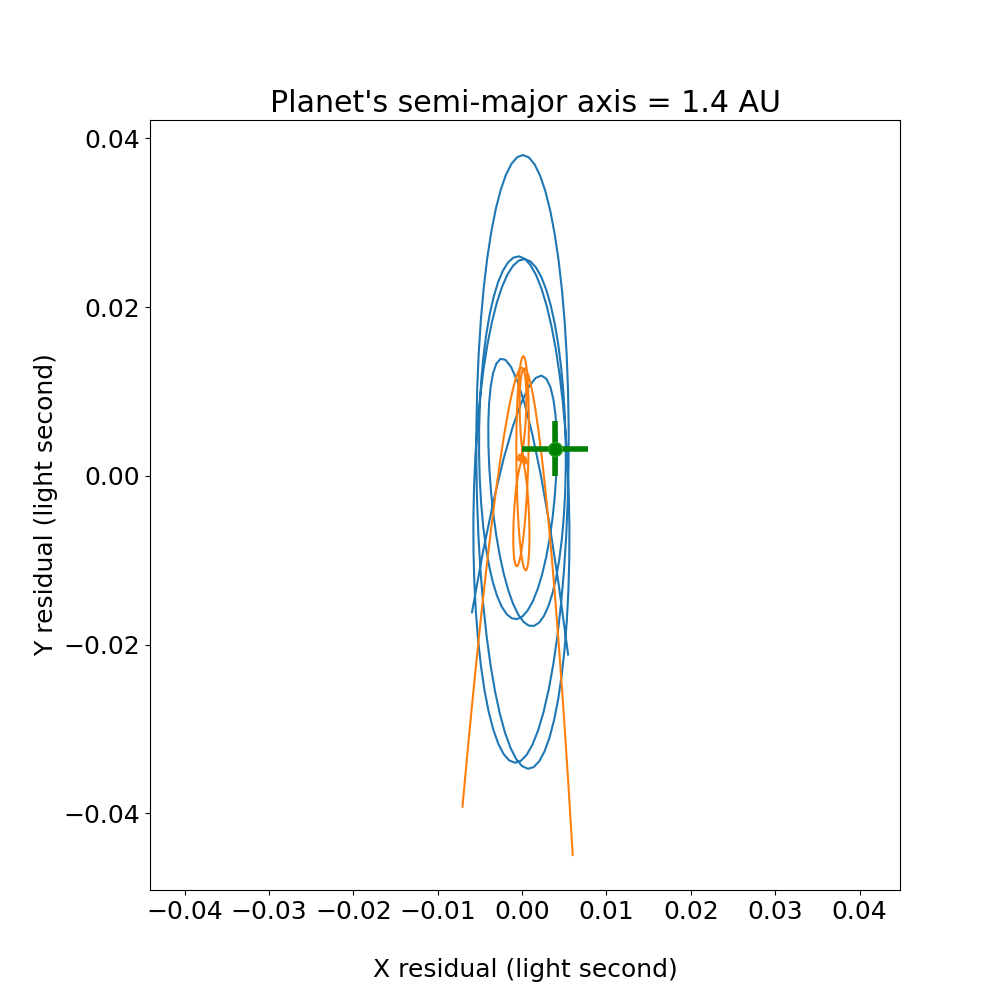}
		\caption{Planet's estimated period = 1.612 yr}
		\label{fig:AU6}
	\end{subfigure}\hfill
	\begin{subfigure}{0.50\linewidth}
		\includegraphics[width=\linewidth]{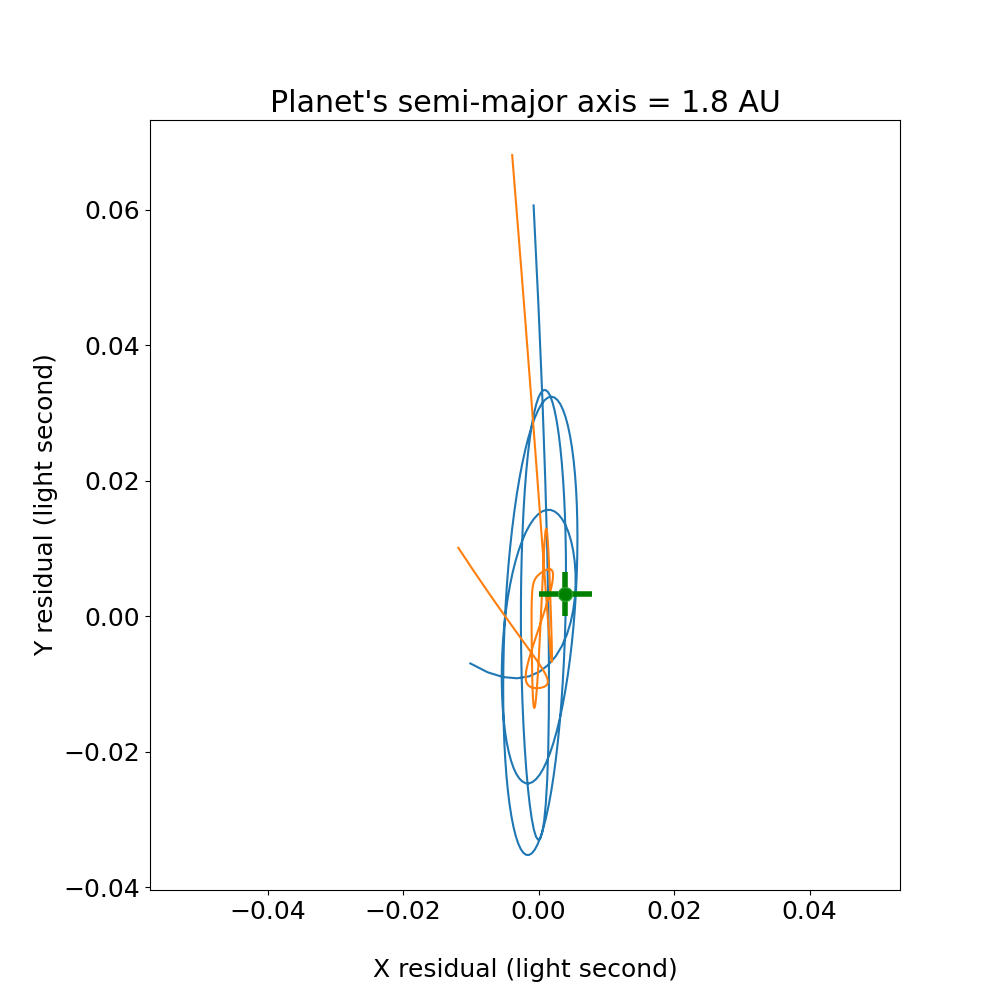}
		\caption{Planet's estimated period = 2.221 yr}
		\label{fig:AU10}
	\end{subfigure}\hfill
	\caption{This figure shows the Habitable orbit estimation of the planet, which can be detected using the setup according to the Fig~\ref{fig:telscpos}. The green error bar is the interferometric error. The mass of the planet taken here is Neptune mass and $6^{th}$ order of polynomial for fitting. The estimated periods of the planet according to their Habitable position are (a) 0.841 years, (b) 1.281 years, (c) 1.612 years, (d) 2.221 years.}
	\label{fig:habitableorbit}
\end{figure*}

We now simulate the measurement of the sky-projected relative position
$(X,Y)$ of \acen~A and B from intensity interferometry at an arbitrary
epoch in 2023, at various SNR levels.  We will then compare the errors
on such a measurement with the astrometric wobble in $(X,Y)$ due to a
planet with different assumed masses.  The distance to the system is
assumed known.

A full analysis pipeline would fit for the masses and initial
conditions as $X(t),Y(t)$ evolve, rather than for a single-epoch
$X,Y$.  From astrometric data alone, the inferred parameters will have
a degeneracy depending on the assumed value of $D$.  Including
kinematic data (not necessarily at very high resolution) would allow
$D$ to be measured as well.  A many-parameter fit combining
astrometric and kinematic data would be rather complicated, but is not
expected to be intractable.  Indeed, an early example of this type of
fit appears already in \cite{1971MNRAS.151..161H}.

In this work, however, our main aim is simply to estimate the
uncertainty in $X$ and $Y$, and how it depends on the SNR.  Thus we do
not include the orbital variation of $X$ and $Y$ in the astrometric
fitting process.  We consider a single-epoch $X,Y$ even though the
observing campaign runs over a long time.

\subsection{The observing configuration}

The setup used is according to Fig.~\ref{fig:telscpos}.  Four
identical telescopes are arranged a few metres from each other. The
diameter of each telescope is $\SI{40}{\centi\metre}$, with a mask width
$b=\SI{2.25}{\centi\metre}$ (see equation~\ref{eq:mask}) oriented at
$104.3^\circ$, so as to approximately match the fringes
at the assumed observing wavelength of $\SI{600}{\nano\metre}$. 

The parameters that set the SNR are as follows.
\begin{equation}
\begin{aligned}
  \Phi & \simeq \SI{1e-4}{{\rm photons}\;\metre^{-2}\,\second^{-1}\,\hertz^{-1}} \\
  A    & \simeq \SI{0.05}{\metre^2} \\
  \Delta t & \simeq \SI{1e-10}{\second}\,.
\end{aligned}
\end{equation}
The spectral photon density $\Phi$ at the observing wavelength follows
from the effective temperatures and angular sizes of the stars (see
Fig.~\ref{fig:phspec}).  The assumed time resolution $\Delta t$
appears realistic and has even been surpassed
\citep[cf.][]{2022AJ....163...92H}.  With these parameters,
equation~\eqref{eqn:truesig} gives
\begin{equation}
  {\rm SNR} \simeq 0.5 \times \gw(\vu)
  \times {\rm channels}^{1/2} \times (\hbox{$t_{\rm obs}$ in s})^{1/2} \,.
\end{equation}
That is to say, if each data point corresponds to $\tav=1\rm\,s$,
noise with $\sigma\simeq2$ would be added to Fig.~\ref{fig:mdata}.

These estimates neglect throughput and detector efficiency. In
real life, these could easily reduce the SNR by a factor of two.
However, such losses could be compensated for by increasing the diameter
of the telescopes to say 60~cm.

Fig.~\ref{fig:simudata} shows some detail of the simulated noisy data.
The Earth's rotation causes the baseline to cross fringes on the
$(u,v)$ plane.  Over $1\,$s the SNR is less than unity.  With $20\,$s
bins a fringe pattern is barely noticeable.  With enough accumulated
data, however, the pattern can be fitted with high precision.

\subsection{Astrometric fitting}

We now apply Bayesian parameter fitting to infer $(X,Y)$ from the
observable (Eq.~\ref{eq:gmeas}) with noise.  Assuming Gaussian noise
with rms $\sigma$, the likelihood will be $\exp(-\frac12\chi^2)$, where
\begin{equation}
  \chi^2 = \sigma^{-2} \sum_k \left( g^{\rm obs}_k
         - \frac{\Delta\tau}{\Delta t} \gw(\vu_k) \right)^2
\end{equation}
and $g^{\rm obs}_k$ is the signal (equation~\ref{eq:gmeas}) at the $k^{th}$ data point for the $u_k$ baseline. $\gw(\vu)$ depends on $X/D,Y/D$ through Equations.~(\ref{eqn:cvisib})
and (\ref{eqn:gbasic}).

There are two distinct time dependencies that go into $\gw(\vu)$.  One
is the fast time variation illustrated in Fig.~\ref{fig:mdata} as the
Earth rotates and $\vu$ changes.  This variation is what allows the
fringes to be measured and hence the astrometry to be inferred.  Then
there is a slow variation of $X$ and $Y$ as indicated in
Fig.~\ref{fig:alphaA} as the stars move in their orbits.

To fit for $X,Y$ we use dynamic nested sampling
\citep{skilling2004nested,skilling2006nested} as implemented in the
      {\em dynesty} code \cite{speagle2020dynesty}.
The normalisation constant $\Delta\tau/\Delta t$ is marginalised out
\citep[see e.g., Eq.~29--33 of][]{10.1093/mnras/stab2391} to avoid
having to fit for it.

As an ambitious but still realistic scenario, consider an observing
program of 1000 nights using 10 spectral channels using the setup of
Fig.~\ref{fig:telscpos}.  This would be time series $10^4$ times what
is illustrated in Fig.~\ref{fig:mdata}.  The stellar positions would
change during this time, necessitating ongoing modifications of the
mask.  In this work we do not attempt to simulate and fit such a large
data set.  Instead we take the signal for one channel and one night
(as in Fig.~\ref{fig:mdata}) and add only 1/100 of the noise level for
one night.  That is, we take the data for one night and reduce the
noise by $({\rm channels} \times {\rm nights})^{1/2}$, compared to
equation~\eqref{eqn:truenoise}.  The total SNR is about $14140$.
Fig.~\ref{fig:estimation} shows the resulting estimation of $(X,Y)$ in
light-seconds.  At $D=1.4\,$pc a light-second is about
$\approx1.5\,$mas.  The 60\% uncertainties along East-West and
North-South are $\approx 5.3 \mu\rm as $ and $\approx 6.2 \mu\rm as $
respectively, which is comparable to the astrometric perturbation due
to an Earth-mass planet $\SI{1}{\rm au}$ from the \acen~A.

As a check that our simplification of simply reducing the noise level
to mimic a longer observing run is reasonable, we have carried out
simulations of one channel up to 35 nights.  The result is shown in
Fig.~\ref{fig:error}, we see that the accuracy indeed improves according to
${\rm nights}^{-1/2}$ as expected.

\subsection{The astrometric wobble}

The observable signal is the difference between the trajectories on
the sky of \acen~A and B, which is to say, the difference between the
two panels of Fig.~\ref{fig:perturbation_plot}.  Over a few years, the
motion is mainly an arc of the 80\thinspace yr binary orbit of
\acen~A-B.  The planetary motion causes a very small wobble, which is barely
visible in Fig.~\ref{fig:perturbation_plot} despite the
unrealistically large planetary mass assumed.  To extract the wobble,
we subtract out a best-fit polynomial

\begin{equation}
\begin{aligned}
 X(t) &= X_0 + X_1 t + X_2 t^2 + X_3 t^3 + X_4 t^4 + X_5 t^5 + X_6 t^6 \\
      &+ \Xwob(t) \\
 Y(t) &= Y_0 + Y_1 t + Y_2 t^2 + Y_3 t^3 + Y_4 t^4 + Y_5 t^5 + Y_6 t^6 \\
      &+ \Ywob(t)
\end{aligned}
\label{eqn:resi-blue}
\end{equation}
leaving a residual $\Xwob(t),\Ywob(t)$, which we call the astrometric
wobble.  The polynomial fit was chosen after some trials as 6-th
degree.  The wobble was further fitted to a sinusoid
\begin{equation}
\begin{aligned}
\Xwob(t) &= X_7 \cos(\omega t) + X_8 \sin(\omega t) + \Delta\Xwob(t) \\
\Ywob(t) &= Y_7 \cos(\omega t) + Y_8 \sin(\omega t) + \Delta\Ywob(t)
\end{aligned}
\label{eqn:resi-orange}
\end{equation}
leaving a smaller residual $\Delta\Xwob(t),\Delta\Ywob(t)$.  The
fitted $\omega$ is the inferred angular frequency of the planet around
its host star.

Fig.~\ref{fig:leastsquare} shows the wobble $\Xwob(t),\Ywob(t)$ and
the residual $\Delta\Xwob(t),\Delta\Ywob(t)$ corresponding to the
orbit in Fig.~\ref{fig:perturbation_plot}.  The blue curve in
Fig.~\ref{fig:leastsquare} is, in effect, the reflex motion of the
host star due to the planetary orbit Fig.~\ref{fig:Binary plot2}, in a
coordinate system that follows the smoothed motion of the host star.
The fitted value of the planet's period is 0.841~yr, which differs
from the input value in Table~\ref{tab:orbitelems} because of orbital
perturbations.  The highly eccentric appearance is partly due to
projection, but is also partly due to the perturbation from \acen~B.

\subsection{Threshold mass and habitable orbit}

Fig.~\ref{fig:massgif} shows the astrometric wobble along with an interferometric error bar (cf.~Fig.~\ref{fig:estimation}).  The wobble residuals (blue and orange) and the interferometric error bar (green) vary according to the planet's masses. 

As the planet's mass is in the range of Jupiter mass or lower,
the perturbed residual is more than the interferometric error. For the mass range 3330--3$\,M_\oplus$ the typical $\Ywob$ is larger than the interferometric error. However, the typical $\Xwob$ becomes comparable to the interferometric error for masses lower than about 16$\,M_\oplus$ ($\approx$Neptune mass). 

Thus the detection threshold for the assumed setup in
Fig.~\ref{fig:telscpos} is roughly a Neptune-mass planet in an
Earth-like orbit.

Studies have been shown that habitable zones exist in the \acen~AB
system (different for both stars).  The range for \acen~A is
(0.9--2.2)~au and for \acen~B is (0.5-1.3)~au as semi-major axis
\cite{2015pthp.confE.101B}.

Fig\ref{fig:habitableorbit} shows the estimation of orbit and period in the habitable zone for \acen~A with Neptune mass of a planet. The variation in blue $\Xwob,\Ywob$ and orange $\Delta\Xwob,\Delta\Ywob$ are according to their increased semi-major axis from \acen~A. As before the green error bars represent the interferometric error.

\subsection{Inferring the host star}

The observable signature of a planet around one of \acen~A or B,
would be that the relative vector between the two stars exhibits an
approximately Keplerian orbit with (say) semi-major axis $a_w$ and
angular frequency $\omega$.  (There will be an uncertainty in $a_w$
because the inclination is not known a~priori, but let us suppose the
inclination is adequately constrained by stability arguments.)  These
observable quantities are given by
\begin{equation}
  a_w = \frac m{M_1} a_p
\label{eqn:astar}
\end{equation}
and
\begin{equation}
  \omega^2 = \frac{G(M_1+m)}{a_p^3}
\label{eqn:orbomega}
\end{equation}
where $m$ denotes the mass of the planet, $M_1$ that of the host star,
and $a_p$ the orbital semi-major axis.  Combining these we have
\begin{equation}
   Gm (1+m/M_1)^{1/3} = (GM_1\omega)^{2/3} \, a_w
\end{equation}
which determines the mass of the planet provided $M_1$ is known.  The
trouble is, while the masses of \acen~A and B are known, it is not
known which of them is $M_1$, which leaves a curious degeneracy in
$m$.

Nevertheless, it seems plausible that the true host star could be
inferred, because of the perturbations of the other star.  The
apparent orbit would show three-body perturbations from Keplerian.
The strength of the perturbations would depend on $a_p/\rtide$ where
$\rtide$ is a characteristic scale for tidal perturbations.  Let us
put
\begin{equation}
   \rtide = (M_1/M_2)^{1/3} \, \rab
\end{equation}
where $\rab$ is the current distance of the other star and $M_2$ is
its mass.  The expression for $\rtide$ is simply the usual Hill radius,
apart from a factor of $3^{1/3}$.  Combining with
equations~\eqref{eqn:astar} and \eqref{eqn:orbomega} and neglecting the tiny
term $m/M_1$, we have
\begin{equation}
   \frac {a_p}{\rtide} = \left(\frac{GM_2}{\omega^2}\right)^{1/3} \frac1{\rab}
\end{equation}
for the amplitude of perturbations.

The above considerations suggest that determining which star is the
host from three-body perturbations would be much harder still than
detecting a planet via astrometry, if indeed a planet is present, but
not hopeless.

\section{DISCUSSION}\label{sec:discussion}

Astrometry to accuracies beyond what is currently available would be a
means to detect or rule out a habitable planet in the \acen\ system.
This paper studies intensity interferometry as a possible means to
this end.  As usual with intensity interferometry, high-end photon
detectors and correlators are required, and accumulating enough SNR
takes a long time.  But set against these is the advantage of a very
simple optical setup, requiring only amateur-grade telescope and no
special infrastructure, because coherence between optical elements is
not required.

\cite{1971MNRAS.151..161H} used intensity interferometry for
astrometry of $\alpha\,$Virginis (Spica).  Recently in
\cite{10.1093/mnras/stab2391} an extension to measure the radii of
both stars together has been suggested.  But this will not work for wide
binary like \acen.  As explained in Section~\ref{sec:fringes} the
requirement that the light buckets are smaller than the fringe width
implies that ${\rm SNR}\propto (R/a)^2$ where $a$ is the separation
between the stars and $R$ is the radius of the larger star.  For Spica
$a/R \sim 10$ whereas for \acen\ $a/R > 10^3$.

To evade this problem, we suggest a sort of matched filter in the form
of a mask placed on the objective.  For large apertures, the mask
could also be placed in secondary optics.

With a declination of $-60^\circ\,50'$ \acen\ is observable year-round
from latitudes further South than $30^\circ\,$S.  The
\SI{1.8}{\metre}-MOA telescope \citep[see
  e.g.,][]{2006apri.meet..272H} at Mt.~John Observatory in New Zealand
can observe \acen\ throughout the year, but at present is not set up
for interferometry.  Hence we consider a different scenario, with
small portable telescopes at a nearby location, but dedicated to
\acen.  \cite{2022AJ....163...92H} have recently demonstrated
intensity interferometry with movable 60\thinspace cm telescopes,
including multi-channel correlation, so a basis for our assumed setup
is already available.  Adding aperture masks with appropriate stripe
width and orientations to do orbital astrometry by
$<\SI{10}{\micro{\rm as}}$ levels, is still a problem to be addressed; 3D printing may be a route to achieve this goal.

\section*{ACKNOWLEDGMENTS}
The authors of this paper greatly acknowledge the Padmanabha cluster, IISER Thiruvananthapuram, India, for high-performance computing time, and thank members of the CTA stellar intensity interferometry working group for discussion.

\section*{Data Availability}
The data underlying this article are available in the article and in its online supplementary material.

\bibliographystyle{mnras}
\bibliography{refs.bib}

\end{document}